\documentclass[12pt]{article}
\usepackage{fullpage}

\usepackage{threeparttable}
\usepackage{lscape}

\usepackage{graphicx}
\usepackage{amsmath}
\usepackage{longtable}
\usepackage[none]{hyphenat}
 \usepackage{setspace}
 \usepackage{url}

\usepackage{lineno}
\usepackage{epstopdf}

 \usepackage{subcaption}

\newcommand{\inlinemaketitle}{{\let\newpage\relax\maketitle}}

\newcommand{\Rebekah}{Rebek\mbox{}ah }

\usepackage{natbib}

\usepackage{textcomp}

\begin{document}

\author{\Rebekah L. Rogers$^1$}

\title{Chromosomal rearrangements as barriers to genetic homogenization between archaic and modern humans}
\date{}

\textbf{Accepted - Molecular Biology and Evolution}

\inlinemaketitle

\begin{center} \Large Research Article \end{center}
\vspace{0.25in}

\noindent 1) Dept of Integrative Biology, University of California, Berkeley \\

\vspace{0.25in}

\noindent \textbf{Running head: } Genome Structure in archaic humans

\vspace{0.25in}


\vspace{0.25in}

\noindent \textbf{Corresponding author:} Rebekah L. Rogers,  Dept. of Integrative Biology University of California, Berkeley, CA 94720 \\
\\
\noindent \textbf{Phone:}  949-824-0614

\noindent \textbf{Fax:}  949-824-2181

\noindent \textbf{Email:} bekah@berkeley.edu

\newpage

\section*{Abstract}
Chromosomal rearrangements, which shuffle DNA throughout the genome, are an important source of divergence across taxa.  Using a paired-end read approach with Illumina sequence data for archaic humans, I identify changes in genome structure that occurred recently in human evolution.  Hundreds of rearrangements indicate genomic trafficking between the sex chromosomes and autosomes, raising the possibility of sex-specific changes.  Additionally, genes adjacent to genome structure changes in Neanderthals are associated with testis-specific expression, consistent with evolutionary theory that new genes commonly form with expression in the testes.  I identify one case of new-gene creation through transposition from the Y chromosome to chromosome 10 that combines the 5' end of the testis-specific gene \emph{Fank1} with previously untranscribed sequence.  This new transcript experienced copy number expansion in archaic genomes, indicating rapid genomic change.  Among rearrangements identified in Neanderthals, 13\% are transposition of selfish genetic elements, while 32\% appear to be ectopic exchange between repeats.  In Denisovan, the pattern is similar but numbers are significantly higher with 18\% of rearrangements reflecting transposition and 40\% ectopic exchange between distantly related repeats.  There is an excess of divergent rearrangements relative to polymorphism in Denisovan, which might result from non-uniform rates of mutation, possibly reflecting a burst of TE activity in the lineage that led to Denisovan.  Finally, loci containing genome structure changes show diminished rates of introgression from Neanderthals into modern humans, consistent with the hypothesis that rearrangements serve as barriers to gene flow during hybridization.    Together, these results suggest that this previously unidentified source of genomic variation has important biological consequences in human evolution.

\clearpage
\section*{Introduction}
Chromosomal rearrangements, which move DNA from one location to another, are a known source of genomic divergence across related taxa.  While distantly related species commonly share large numbers of orthologous genes \citep{Jaillon2004,Putnam2008,Murphy2005}, syntenic tracts display genome shuffling across all metazoan clades \citep{Jaillon2004,Putnam2008,Murphy2005, Eichler2003,Bhutkar2008, Bennetzen2000}.  Such genome shuffling events are a source of genetic novelty that can form new genes \citep{Miller1993,Bennetzen2005}, modify expression patterns \citep{Duhl1994}, and create linkage between previously unlinked genes \citep{Rieseberg2001}.  Mammalian genomes experienced active genome shuffling \citep{Murphy2005}, and even close relatives such as humans and gibbons differ by over a hundred syntenic breaks \citep{Roberto2007}.  This alternative source of genomic variation remains understudied in comparison to SNPs, especially in human evolution.

Yet, these large-scale mutations that alter whole genomic segments can affect gene expression and function.  Genome shuffling and transposition can modify expression patterns for neighboring genes \citep{Slotkin2007,Kidwell1997}, and positionally relocated genes produce regulatory changes between humans and chimpanzees  \citep{De2009}.  Movement of DNA, especially when facilitated by transposable elements, can fortuitously place regulatory machinery next to genes, thereby modifying gene expression \citep{Kidwell1997, Cridland2015} or chromatin modeling effects \citep{Argeson1996,Michaud1994,Slotkin2007}.   As with all variation, while some of these mutations may be beneficial, many are associated with disease.  Chromosomal rearrangements are associated with multiple types of cancer \citep{Mitelman2007}, infertility \citep{Stern1999,Dong2012,Alves2002}, spontaneous abortions \citep{DeBraekeleer1990}, autism \citep{Folstein2001}, and language disorders \citep{Lai2000,Tomblin2009}.   Thus, a better understanding of how rearrangements accumulate along the human lineage will have direct impacts on human health. 

 Advances in next-generation sequencing allow whole genome surveys of two close relatives of modern humans: Denisovans and Neanderthals.  High coverage, high quality, low contamination sequence data is available for one individual from each archaic group \citep{Meyer2012,Prufer2014}.  SNP diversity in Denisovans and Neanderthals has produced a clear snapshot of modern and archaic human differences as well as human-Neanderthal interactions.    Archaic and modern humans diverged 270,000-440,000 years ago \citep{Reich2010} and spread through Eurasia during independent migration events \citep{Sankararaman2012}.  Humans and Neanderthals coexisted with overlapping ranges for tens of thousands of years, interbreeding  with archaic humans around 47,000-63,000 years ago \citep{Sankararaman2012}.   The average Eurasian typically carries $\sim$2\% of archaic human DNA \citep{Sankararaman2014,Green2010,Prufer2014} and understanding the mutations that differentiate modern and archaic humans will clarify mutations that might have origins in archaic groups.  The availability of  closely related outgroups for modern humans additionally offers the unique opportunity to trace recent changes that have appeared in modern human genomes. 

 The Denisovan genome has been sequenced to 38X \citep{Meyer2012} and the Neanderthal genome was sequenced to 52X \citep{Prufer2014} allowing an accurate portrait of genomic variation between archaic and modern humans.   I have assayed this high-quality genome sequencing data for recent changes in genome architecture that differentiate the archaic and modern human reference genomes.  In Illumina sequencing data, chromosomal rearrangements are manifest in cases where two reads for a single read pair map to divergent chromosomal locations \citep{Cridland2010}.  Using paired-end read mapping data, I identify hundreds of changes in genome structure between archaic humans and the modern human reference genome.  Such methods have successfully identified genome structure changes in model organisms with high accuracy using Illumina data \citep{Corbett2012, Rogers2014,Cridland2010} and a similar paired-end approach identified structural variants in modern human genomes using 454 sequence data \citep{Korbel2007}.  Thus, these methods now can be applied to archaic samples with high confidence.  DNA fragmentation during degradation produces short insert sizes in Illumina libraries for archaic humans \citep{Meyer2012,Prufer2014}, and only a fraction of alignments are useful for surveys of genome structure.  These data offer a limited portrait of genome structure variation between modern and archaic humans, allowing us to identify this previously unanalyzed source of genomic variation.
 
 Previous CNV detection in Neanderthals has focused on coverage changes to identify dozens of duplications in modern and archaic humans \citep{Prufer2014}.  Coverage-based assays detect many CNVs with high validation rates \citep{Sudmant2010,Alkan2009}, but they may overestimate the number of independent duplication events if duplicates experience secondary modification or if rearrangements are complex \citep{Rogers2014}.  Furthermore, new genomic locations of copy number variants cannot be localized using coverage alone.  In contrast, paired-end read mapping offers additional information that can identify duplicative changes as well as non-duplicative changes in genome structure.  Read pairs can therefore resolve complex rearrangements with greater precision than coverage based assays and will not be limited by the size of the translocated sequence, up to the length of sequencing reads. Major genome sequencing projects in modern humans have generated human cell lines \citep{ThousandGenomes,HGDP}, which are prone to genomic rearrangements unrelated to natural variation.   Here, archaic genomes collected without (the possibility of) generating cell lines offer one major advantage as they will be free of artificially induced rearrangements.  Thus, the data presented here include newly-identified variants representative of natural variation including hundreds of mutations not found in previous assays.
 
 Based on these newly identified chromosomal rearrangements, genes adjacent to genome structure changes in Neanderthals are associated with testes specific expression in modern humans, consistent with evolutionary theory that new genes commonly form with expression in the testes \citep{Betran2002,Kaessmann2010,Assis2013}.  Multiple cases of genomic trafficking between the autosomes and the sex chromosomes differentiate modern and archaic humans, raising the possibility of sex specific changes in human evolution.  There is an excess of divergent mutations in the Denisovan genome, possibly driven by a burst of TE activity.  Further, loci containing genome structure changes show diminished likelihood of introgression, consistent with the hypothesis that genome structure changes serve as one potential barrier to genetic homogenization between modern and archaic humans through negative selection after interbreeding.  Finally, in a chimeric construct formed through chromosomal rearrangement at the \emph{Fank1} locus, a sperm specific promoter is combined with a previously untranscribed region to create a new exon.  Subsequent duplication of the newly formed gene sequence in archaic humans points to exceptionally rapid evolution in genome structure at the \emph{Fank1} locus.  Together, these results suggest that chromosomal rearrangements are a common source of variation between modern and archaic humans capable of influencing human biology and evolution.

\section*{Results}
Here, I identify genome structure changes between modern and archaic humans.   I describe patterns of genome structure changes on the autosomes and sex chromosomes, expression patterns of neighboring genes, and likelihood of introgression into modern humans.  Finally, I describe new gene formation through chromosomal rearrangement with rapid changes in copy number in modern and archaic humans.  

\subsection*{Genome Structure Variation}
  Chromosomal rearrangements identified here include mutations dispersed across chromosomes and those that moved DNA within a single chromosome over a distance greater than 1 Mb.  Such variants are expected to capture translocation, dispersed duplication, gene conversion, ectopic recombination, retrogene formation and transposition by selfish genetic elements, all molecular mechanisms that move DNA from one genomic location to another.  These mutations are unpolarized with respect to the ancestral state but reflect sites where synteny is different between Neanderthals and the modern human reference genome.  Paired-end mapping from Neanderthal sequence data identifies 985 changes in genome structure while data from Denisovan indicates 1330 changes in genome structure (Table \ref{Mutations}, \ref{ByChrom}).  A total of 326 variants are identified with at least 3 read-pairs and less than 100 read pairs  both in the Denisovan genome and in the Neanderthal genome sequence.  

\begin{figure}
\includegraphics[scale=0.65]{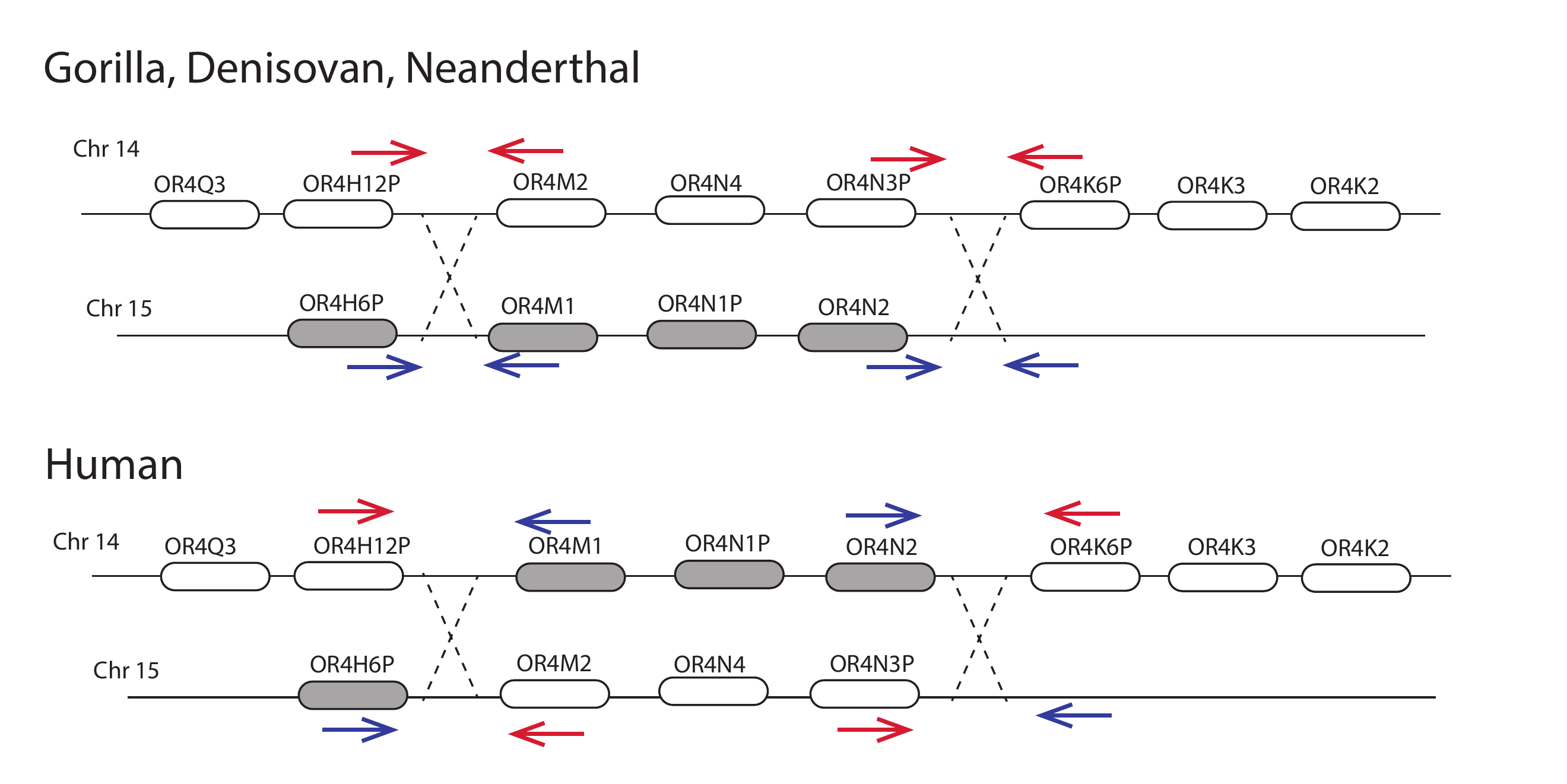}
\caption{\label{Olfactory} Ectopic recombination captures olfactory receptors in the human lineage. I identified breakpoints of a change in genome structure using abnormal paired-end read mapping to chromosomes 14 and 15 in Illumina short read data for the high coverage Neanderthal and Denisova genomes.  Alignments with the outgroup genome match with the state inferred for archaic humans based on paired-end reads from archaic humans. } 
\end{figure}

Paired-end read mapping can identify changes in genome structure that occurred between archaic and modern humans, but in isolation cannot identify in which lineage the mutations occurred.  Polarizing mutations against outgroups, I identify 348 variants in Neanderthal and 357 variants in Denisova where archaic genomes carry the ancestral rather than the derived state.  One well defined example, is shown in Figure \ref{Olfactory}.  Abnormal read pair mapping indicates a reciprocal rearrangement affecting chromosome 14 and chromosome 15 in regions containing multiple olfactory receptors.  Read pair mapping indicates two breakpoints capturing 64.1 kb on chromosome 14 and 65.2 kb on chromosome 15.  Sequences that correspond to abnormally mapping read pairs in human reference genome match to the same locations in gorilla, independent confirmation of the rearrangement state identified in archaic genomes (Figure \ref{Olfactory}).   For X and autosomal rearrangements where Neanderthals (rather than the human reference) carry the derived state 212/287  rearrangements have 1000 bp on one side of at least one breakpoint that exhibits coverage 2 standard deviations above the mean coverage, suggestive of predominantly duplicative rearrangements.  In Denisovan the proportion is slightly lower with 327/539 derived variants on the X and autosomes displaying increases in coverage.  Secondary mutations may exaggerate the instance of duplicative rearrangements and these numbers represent an upper bound on the number of duplicative changes.    
  
The proportion of derived vs. ancestral rearrangements along the archaic lineage is significantly greater in Denisovan than Neanderthal  ($\chi^2 = 11.6917$, $df = 1$, $p-value = 0.0006278$) and more rearrangements are seen in Denisovan than Neanderthal.  Yet, data from the two archaic humans demonstrate agreement in the ability to identify rearrangement mutations that occurred in the modern human genome (348 using Neanderthal and 357 using Denisovan), an indication that these differences are unlikely to be driven by higher false negative rates in Neanderthals.   A total of 161 genes have transcription start or stop sequences within 10 kb of changes in genome structure in Neanderthal and 222 genes lie adjacent to genome structure changes in Denisovans.  In Denisovan these genes are associated with gene ontology categories of keratin, flotillins and caveola, microtubules, and fibronectins, suggesting an association with structural peptides (Table \ref{GO}).  In Neanderthal, which has fewer rearrangements, only an association with flotillins and caveola is significant at an EASE cutoff of 1.0 (Table \ref{GO}).   

\subsection*{Rearrangements and recombination}
\begin{figure}
\begin{subfigure}{.5\textwidth}

\includegraphics[scale=0.47]{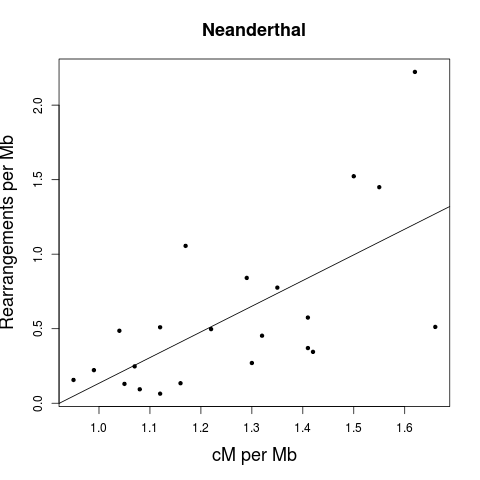}
\caption{}
\end{subfigure}
\begin{subfigure}{.5\textwidth}
\caption{}
\includegraphics[scale=0.47]{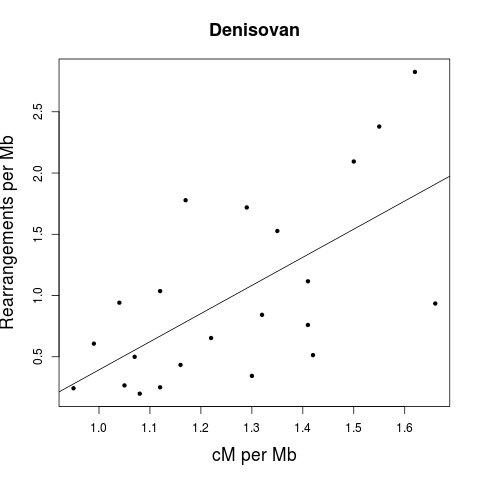}
\caption{}
\end{subfigure}
\caption{\label{RecombRate} Incidence of rearrangements identified in (A) Neanderthal and (B) Denisova vs recombination rate by chromosome, for all autosomes.   Both samples show a significant positive correlation between incidence of rearrangements and recombination rates (Neanderthal $R^2=0.402$, $P=0.0009$; Denisovan $R^2=0.376$, $P=0.001433$).  Higher rates of rearrangements per basepair are observed on more highly recombining chromosomes.}
\end{figure}

The rate of rearrangements per basepair is inversely correlated with chromosome size in Neanderthal ($R^2=0.24$, $P=0.0088$) and Denisovan ($R^2=0.20$, $P=0.016$, Figure \ref{RearrLength}), challenging the hypothesis that rates of formation are uniform across the genome.  The correlation becomes insignificant when considering chromosome length in centimorgans \citep[data from][]{Venter2001} and the coefficient of variation is considerably lower (Neanderthal $R^2=0.036$, $P=0.195$; Denisovan $R^2=0.01$, $P=0.28$, Figure \ref{RecombLen}).  The differences in results when considering physical vs recombinational length imply that the correlation is likely to be driven by recombination rate differences across chromosomes.   Furthermore, there is a strong correlation between rearrangements per bp and recombination rates (Neanderthal $R^2=0.402$, $P=0.0009$; Denisovan $R^2=0.376$, $P=0.001433$, Figure \ref{RecombRate}).  Power to detect genome architecture changes depends heavily on coverage of paired-end reads (Rogers et al. 2014).  There is no disparity in coverage of properly paired reads across chromosomes that could explain the increased number of rearrangements observed on smaller, more highly recombining chromosomes (Table \ref{PairCov}).  Hence, it is unlikely that the observed heterogeneity in the number of rearrangements across the genome is a methodological artifact related to heterogeneous coverage.

\subsection*{Repetitive Elements}
 Repetitive sequence locations provided by the UCSC genome browser for hg19/GRCh37 (downloaded June 2015) suggest that 315 rearrangement calls in Neanderthals (32\%) have both breakpoints in repetitive elements.  These loci could represent ectopic recombination facilitated by TEs, smaller-scale gene conversion acting to homogenize TE sequences, or nested TEs at transposition hotspots.   An additional 131 rearrangements identified using Neanderthals (13\%) have one breakpoint in a transposable element, representing novel TE insertions.  In Denisovan, the pattern is similar but numbers are higher, with 543 (40\%) having both breakpoints in TEs, and 236 (18\%) with only one breakpoint in a TE sequence.  While the higher numbers of rearrangements with both breakpoints in TEs might be affected by higher error rates in Denisovan, the number of transposition events is not expected solely from error prone reads.  The greater association with repetitive sequences in Denisovan is highly significant ($\chi^2=50.24$, $df=1$, $P=1.829\times10^{-9}$) raising the possibility that TEs may have been more active in the lineage leading to Denisovan than that leading to Neanderthals.  Denisovan and Neanderthals carry roughly the same number of non-repeat rearrangements, with 539 non-TE rearrangements in Neanderthal and 551 in Denisovan. Thus, the excess of rearrangements observed in Denisovan is likely to be due to repetitive element mediated DNA movement, both through active transposition and through passive effects of facilitating ectopic recombination.  
 
\subsection*{Genomic trafficking and sex chromosomes}
\begin{figure}
\includegraphics{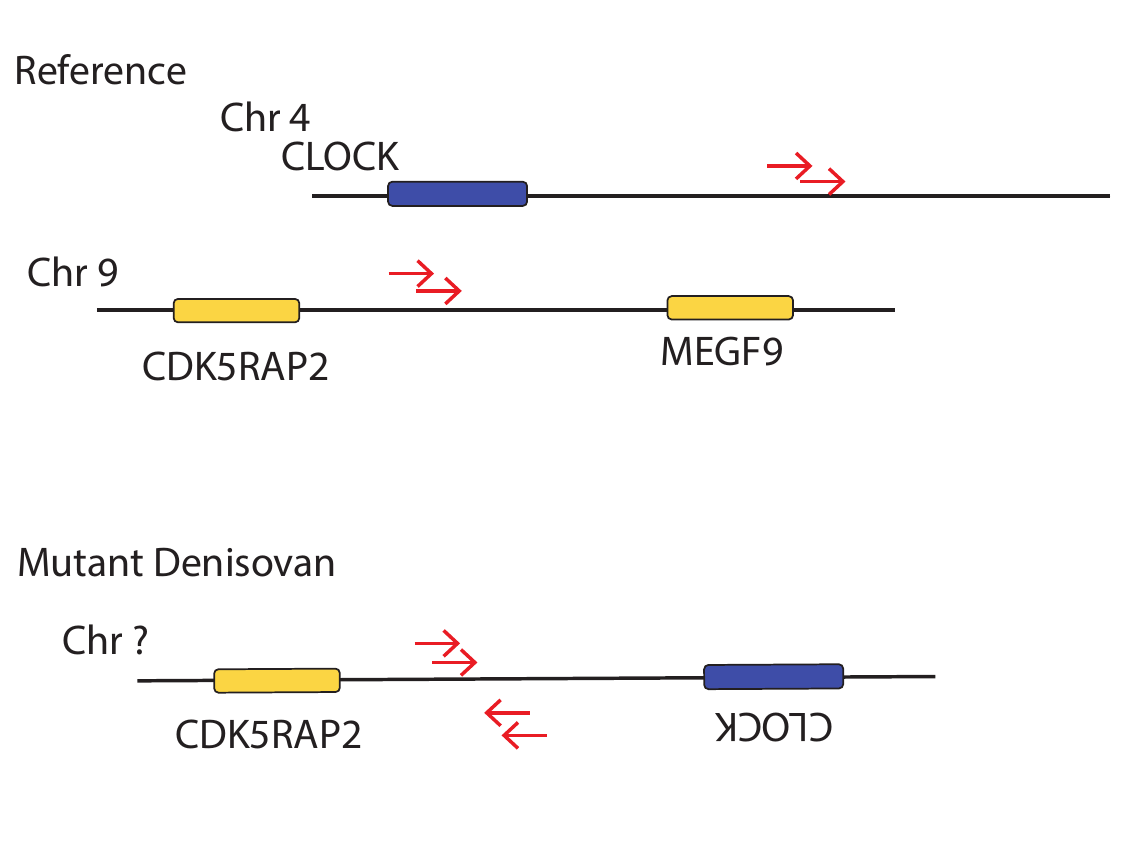}
\caption{\label{Circadian} Change in genome structure flanking \emph{CDK5RAP2}.  The current chromosomal location cannot be inferred from limited data in archaic humans and the mutation might reside on chromosome 4 or chromosome 9.  Nonsense mutations in \emph{CDKRAP2} produce pathogenic microcephaly while \emph{CLOCK} regulates circadian rhythm in model organisms.  Both genes are known to be expressed in neurons. }
\end{figure}

Both high coverage genomes for archaic humans sampled to date are female specimens, as confirmed by low Y coverage relative to the X and autosomes \citep{Prufer2014, Meyer2012}.  Using female archaic samples, I am able to identify a total of 158  translocations from the autosomes to the Y in modern humans or from the Y to the autosomes in archaic humans, 110 using Denisova and 98 using Neanderthal.  Of these variants, 3 in Denisova and 4 in Neanderthal are between the X and Y, in regions outside the pseudoautosomal regions PAR1, PAR2, and XTR/PAR3.  With autosome-autosome rearrangements, paired-end reads are unable to identify the direction of rearrangement even when it is clear that rearrangement has occurred (e.g. Figure \ref{Circadian}), especially in cases where outgroup genomes are poorly assembled or uninformative due to secondary mutations.   However, with female genomes, the presence of a full Y is excluded, demonstrating that these rearrangements currently reside on the autosomes or the X.  Using the chimpanzee as an outgroup genome, ancestral state for Y-autosome translocations was identified to determine the direction of DNA movement. I can identify 14 cases where there is a match for both reads in a pair within 1 kb of one another on an autosome in chimpanzee, clearly indicating movement to the Y in modern humans.   In contrast there are 30 clear cut cases where one read in the pair maps to the chimpanzee Y, indicating Y to autosome movement in archaic humans.  The ancestral state for the remaining Y variants cannot be established, possibly due to limitations of chimpanzee genome assemblies.  The X chromosome does not contain an excess of rearrangements per bp in comparison with its size (Figure \ref{RearrLength}). However, the Y chromosome appears to contain many rearrangements relative to its size (Figure \ref{RearrLength}), especially when one considers the inability to identify movement to the Y chromosome in archaics and from the Y chromosome in modern humans given these female samples.  

\subsection*{Polymorphism for genome structure variants in modern humans}
Ten samples of Illumina sequence for modern humans confirm genome structure variants identified in archaic human genomes.  Data from modern humans validate 556 genome structure variants that were identified in Neanderthal and 548 variants that were identified in Denisovan.  Such agreement indicates that the observed excess of derived mutations in Denisovan is due to Denisovan-specific mutations rather than an excess of derived ancient polymorphism.  The bioinformatic methods implemented here cannot determine whether mutations are heterozygous or homozygous, especially for young, non-duplicative rearrangements with little sequence divergence across copies and these presence-absence spectra offer indirect estimates of rearrangement frequencies.   In Neanderthal there are 139 (25.0\%) ascertained at a sample frequency of 1/10 and 116 (20.9\%) at a sample frequency of 10/10.  In Denisovan, there are 147 (26.8\%) found at a sample frequency of 1/10 and 104 (18.9\%) identified at a frequency of 10/10 in modern humans.  The human reference genome lacks these mutations, and thus many of these mutations will be segregating at high frequency in modern humans.  Folded presence-absence spectra for mutations identified in archaic genomes show large numbers of rare or common variants, with fewer moderate frequency variants (Figure \ref{SFS}).   Mean frequency of genome structure variants identified in Neanderthal sequences and confirmed using modern human genomes is 4.8/10, and mean frequency of genome structure events identified in Denisova sequences is also 4.8/10.   Fifty-two Y variants in Neanderthal and 54 Y variants in Denisova can be confirmed using modern human genome sequences, suggesting that these are unlikely to be artifacts of DNA damage or preparation methods specific to sequencing of ancient DNA.    

To confirm that high frequency variants observed in modern humans are not driven by bioinformatic artifacts, I validated the 116 rearrangements observed using Neanderthal sequencing data using PacBio long molecule sequences available from a haploid complete hydatidiform mole provided by Pacific Biosciences (\url{http://datasets.pacb.com/2014/Human54x/fast.html}, Accessed March 2015).  In this haploid modern human sample, I can confirm 99/116 high frequency rearrangements for a validation rate of 85\%.  Given that this PacBio data is taken from a different human sample that will have different segregating rearrangements, the validation rate is very high and considering allele frequency expectations is roughly in line with the 96\% validation rate observed in less repetitive model organisms \citep{Rogers2014,Cridland2010}.  There is no significant difference in confirmation rates for ancestral (12/14)  vs derived (32/42) mutations identified in Neanderthals ($\chi^2=0.00076$, $df=1$, $P=0.978$).

To determine whether rearrangement mutations are accumulating under constant neutral processes, one can compare polymorphism to divergence for rearrangements and for neutral intergenic SNPs processed according to similar criteria (see Methods).  Among 887 X and autosomal rearrangements identified using Neanderthals, 504 are currently polymorphic in modern humans while 383 are divergent between humans and Neanderthals.  Divergence equals 80\% of polymorphism, with no significant difference between divergence rates for rearrangements and neutral SNPs (Table \ref{Divergence}; $\chi^2=0.692$, $df=1$,$P=0.4055$).  However, for Denisovan, there are 725 divergent rearrangements identified compared to 494 polymorphic X and autosomal rearrangements, a significant departure from results in Neanderthals ($\chi^2= 54.0326$, $df=1$, $P=1.972\times10^{-13}$) and for neutral intergenic SNPs ($\chi^2=70.08$, $df=1$, $P=2\times10^{-16}$).  Rearrangements, especially those related to transposable elements, may accumulate according to non-constant dynamics, with rate heterogeneity as `bursts' of TE activity occur at discrete timepoints.  Such heterogeneous mutation rates violate the assumptions of McDonald-Kreitman type tests.  Thus, the excess of divergent rearrangements identified in Denisovan could be the product of demographic effects resulting in accumulation of rearrangements in comparison with SNPs, positive selection on rearrangements, or a burst of TE-related activity in the ancestor of Denisovans, effectively decoupling mutation rates from current segregating polymorphism.  Given the excess of TE associated rearrangements in Denisovan, it seems likely that non-uniform mutation rates are a major contributing factor to the observed excess of divergence.

\subsection*{Testes-specific genes}

In model organisms, new genes commonly appear with expression in the testes \citep{Betran2002,Kaessmann2010,Assis2013}, and testes-expressed genes show evidence of rapid evolution \citep{Wyckoff2000,Haerty2007,Voolstra2007,Dorus2010}.  To determine whether genome structure changes identified in archaic humans are associated with testes expression, I analyzed two independent sources of gene expression data for modern humans.  Among genes associated with chromosomal rearrangements I identify 124 genes within 10kb in Neanderthals that could be assessed for expression using the Human Protein Atlas \citep{Fagerberg2014} (\url{http://www.proteinatlas.org/}, accessed Oct 2014).  Of these genes, 15 are associated with testis-specific expression, an overrepresentation compared to expectations based on random resampling ($P=0.0084$, Table \ref{Testes}).  Additionally, using previously published data on divergence in gene expression between humans and chimpanzees, 9 out of 25 genes associated with genome structure that could be assayed show gene expression changes only in the testes (Table \ref{Testes}).    The expression divergence data for humans and chimpanzees survey a limited number of genes, but it offers independent confirmation that changes in genome structure identified in Neanderthals are associated with testes-specific effects.  Denisovans, in contrast, show a different pattern.  Chromosomal rearrangements are associated with testis-specific expression in only 14 genes out of 182 ($P=0.22$) and 11 out of 46 genes showing testes-specific gene expression changes between humans and chimpanzees ($P=0.24$, Table \ref{Testes}).  These results offer two independent confirmations that genes adjacent to rearrangements in Neanderthals but not Denisovans are associated with testes-specific effects.   

\subsection*{Loci with chromosomal rearrangements are resistant to gene flow}
Neanderthals and modern humans interbred in Eurasia after the human migration out of Africa and the average European shares roughly 2\% of their genome with Neanderthals \citep{Sankararaman2014,Prufer2014,Green2010}.  Some regions of the genome are more prone to introgression than others, and gene content, gene expression, and sex chromosome status influence introgression rates \citep{Sankararaman2014}.  A newly analyzed introgression dataset from Steinr\"ucken et al.  (\url{http://dical-admix.sf.net}) places posterior probabilities on introgression for each locus in the genome, offering more nuanced information.   Mean probability of introgression genome wide is 0.012 ($\sigma=0.00083$), while chromosomal rearrangements experience mean introgression probability of 0.008 ($P=0.0015$, Table \ref{Introgression}). Many chromosomal rearrangements are detrimental, and it is possible that selection against new mutations could reduce introgression rates.  However, when I consider only cases where Neanderthal holds the ancestral state whereas the modern humans hold the derived state, excluding the possibility of selection against newly formed detrimental mutations, mean probability of introgression is 0.007, a significant reduction from neutral expectations ($P=0.0015$, Table \ref{Introgression}).  Independently analyzed introgression calls from Sankararaman et al. 2014, also point to reduced likelihood of introgression at regions containing genome structure changes.  When Neanderthal carries the ancestral state,  228 out of 748 (30\%) regions experience introgression in at least one sampled haplotype. When Neanderthal holds the derived rather than the ancestral rearrangement state, 181 out of 844 (21\%) regions experience introgression.  Results from a third study of introgression suggest more extreme introgression rates of only 2.2\% for ancestral mutations and 2.7\% for derived mutations compared to background rates of 19.1\% ($P<10^{-16}$) \citep{Vernot2014}.  These observed proportions of sites associated with introgression display a significant departure from genome-wide background introgression rates of 35.64\% for both derived and ancestral rearrangements ($P<10^{-16}$, $P\leq0.04$ respectively).  

\subsection*{Formation of a new gene expressed in the testes}
Among genes affected by chromosomal rearrangements, one shows signs of dynamic changes in genome structure.  \emph{Fank1} is a testis-specific gene in modern humans, and its ortholog functions during the transition from diploid to haploid chromosome number during meiosis in mice \citep{Zheng2007}.   $d_N/d_S$ for this gene is high, suggesting rapid evolution in the human lineage ($d_N/d_S=1.2$, $\chi^2=20.494$, $df=2$, $P=3.5\times10^{-5}$). The first exon of \emph{Fank1} is flanked by 6 different sets of read-pairs indicating rearrangement between chromosome 10 and the Y in Neanderthal and in Denisovan.  Coverage for both archaic genomes is consistent with 6 duplications of the first exon of \emph{Fank1} although coverage across the multicopy region varies (Figure \ref{FankDepth}-\ref{DeniFankDepth}).  The region displays high heterozygosity in both the Neanderthal and Denisovan genome, which correlates well with coverage (Figure \ref{FankHet}), an indication of diverged paralogs.

\begin{figure}

\includegraphics[scale=0.21]{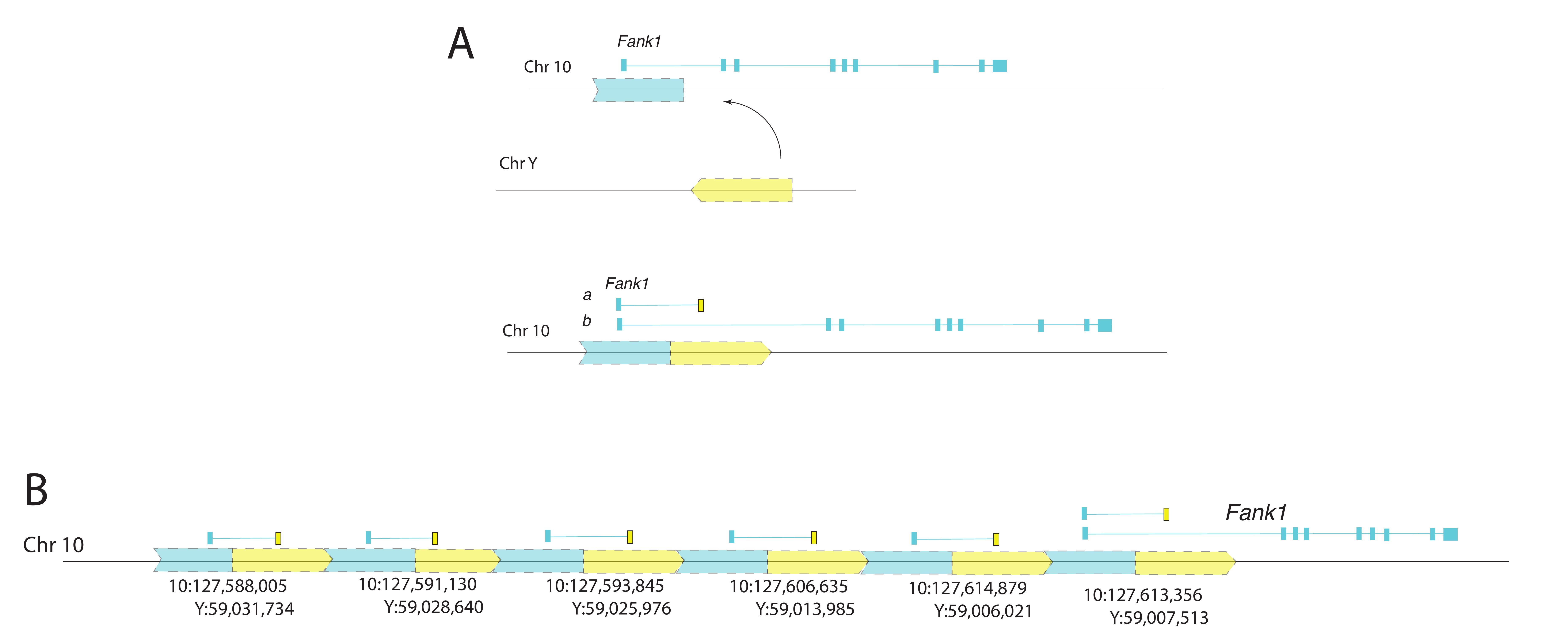}
\caption{\label{FankStruct}Origins of a newly transcribed sequence due to translocation at the \emph{Fank1} locus.  A) A chromosomal rearrangement moved a region of the Y to the region adjacent to the first exon of \emph{Fank1}.  Paired end reads in RNA-seq data from testis of a modern human indicate fusion transcripts uniting the first exon of \emph{Fank1} and an unannotated region that matches with the Y.  The translocation is present in all modern humans surveyed and in the Neanderthal and Denisovan genome sequences.  Downstream exons of \emph{Fank1} outside the region with the rearrangement are still transcribed in the testes.  B)  Structure of copy number variation for the \emph{Fank1} locus in Neanderthal.  Breakpoints indicated by abnormally mapping read-pairs are labeled.  Duplication of the rearranged segment at the \emph{Fank1} locus resulted in roughly 6-fold copy number variation for the first exon of \emph{Fank1} and a newly transcribed exon derived from a segment formerly located on the Y.  Exact order of copies with specific breakpoints within the 6X cassette is not known, though independent breakpoints are indicated in the paired-end read mapping and coverage data.  The Denisovan genome sequence data shows a similar 6-fold expansion of the region, however only some of the copies share precise breakpoints.  } 

\end{figure}

Figure \ref{FankStruct} provides one genetic structure produced by rearrangement, secondary tandem duplication, and deletion that can explain the observed coverage variation and abnormally mapping read-pairs in archaic genome sequences.   The Y contains a segment of DNA that experienced rearrangement through ectopic recombination or transposition.  The mutation  moved a segment of DNA from the Y at $\sim$58.97-59.03 Mb over to chromosome 10, inserting the sequence at $\sim$127.61 Mb (Figure \ref{FankStruct}A).   Subsequent expansion via tandem duplication then created copy number variation, with 6 copies in archaic genomes (Figure \ref{FankStruct}B).  Multiple partial duplications during copy number expansion or secondary deletions \citep[e.g.][]{Rogers2014} result in multiple unique breakpoints in the paired-end read data.  The exact ordering of individual copies in the region cannot be determined using Illumina sequencing, but one  structure is shown in Figure \ref{FankStruct}B.  In modern human DNA sequences that are not affected by the same degradation and damage as archaic genomes, both paired-end read mapping information and coverage confirm that the region is subject to rearrangements in modern humans, with variation in copy number (Figure \ref{HGDPCov}-\ref{HGDPYCov}).   Additionally, split read mapping of long molecule sequences from a haploid genome publicly available from PacBio (\url{http://datasets.pacb.com/2014/Human54x/fast.html}, Accessed March 2015) confirms rearrangement between the Y and chromosome 10 at the first exon of \emph{Fank1}.  Previous paired end 454 sequencing of restriction enzyme fragments has identified a translocation between the Y chromosome and the \emph{Fank1} locus as well \citep{Chen2008}.    

To determine whether the duplicated first exon of \emph{Fank1} can drive expression of adjacent sequence, I obtained testes expression data from the ENCODE project (\url{www.encodeproject.org}, from Michael Snyder's lab). I identify a total of 4 read-pairs in the transcriptome data that indicate fusion transcripts of sequence on chromosome 10:127588640-127600391 and Y:59009858-59031127, each read mapping with 101 matches and no mismatches.    Read-pairs are located with the orientation expected based on the orientation of the Y-autosome translocation.  Coverage in the RNA seq data from Y:59020171-59031127 indicates that this promoter can drive expression of the relocated region from the Y, thereby forming a novel gene sequence.  The new gene would not carry either of the ankyrin conserved domains from \emph{Fank1}, as they are not found in the first exon, unless there exists an unidentified fusion transcript incorporating the new exon into the \emph{Fank1} mRNA.  The standard isoform of \emph{Fank1} (ENST00000368695) is expressed with an FPKM of 25.6 across all exons, and is not truncated by the rearrangement (Figure \ref{FankStruct}).

\section*{Discussion}

\subsection*{Genome structure changes in hominids}

Paired end read mapping identifies 985 chromosomal rearrangements using Neanderthal genome sequences and 1330 using Denisovan sequences.   Modern human genomes validate 556 genome structure variants that were identified in Neanderthal and 548 variants that were identified in Denisovan.  This validation rate is extremely high given that SFS are generally skewed towards rare variants.  Additionally, 99/116 variants identified in Neanderthal and ascertained at high frequency are validated by PacBio long molecule sequencing data.  Furthermore, 348 rearrangements in Neanderthal and 357 in Denisovan match with the  ancestral state, indicating mutations occurring in modern humans.   Large numbers of genome shuffling events contribute to the divergence between archaic and modern humans, with higher rates of genome shuffling in Denisovan in comparison to Neanderthal.   A greater association with repetitive elements in Denisovan as well as high rates of divergence for Denisovan rearrangements suggest a burst of selfish genetic element movement in the Denisovan lineage.
 
 The number of rearrangements per base pair shows an inverse correlation with chromosome size in Neanderthal and Denisovans.  In mammals, chromosomes experience a minimum of one recombination event per meiosis \citep{Darlington1937} and chromosome size  correlates with recombination rates in model organisms and in humans \citep{kaback1996,Lander2001}.  Further, the number of rearrangements per base pair correllates well with recombination rates, suggesting that recombination plays a major role in the generating chromosomal rearrangement differences between modern and archaic humans.  If selection were removing variation, one would expect selection purging (largely detrimental) variation to be weaker with low recombination resulting in the accumulation on larger chromosomes with lower recombination rates compared to smaller chromosomes with higher recombination rates.   However, I observe more rearrangements on more highly recombining chromosomes, in contrast to what one expects if patterns were driven by negative selection.  Thus, one would not expect that selection would produce the particular trend observed.  Together, these results imply that genomic location can influence lability of gene sequences, and that mutational pressures will be higher on smaller, more highly recombining chromosomes.

\subsection*{Gene shuffling}
One case of shuffling affecting olfactory receptors is particularly well-resolved with two breakpoints of the rearrangement identified and with clear agreement between archaic humans and outgroup genomes, pointing to human-specific mutation.  Previous work using microarrays has successfully identified copy number variation for olfactory genes \citep{Hasin2008} but with next-generation sequencing  I can identify shuffled loci even when there is no corresponding change in copy number.  Mammalian olfactory receptors fall into distinct clades that show signatures of shuffling and rearrangement across the genome \citep{Niimura2003} and ectopic recombination events are common in regions with olfactory receptors \citep{Trask1998,Giglio2001}.  Evolution of olfactory receptors has been subject to strong selection along the mammalian lineage and there are signatures of positive selection on olfactory genes in humans \citep{Clark2003}.  The observed changes in olfactory receptors may therefore be due to adaptive mutations, permissive shuffling due to mutational pressures, or novel detrimental mutation ultimately destined for loss.  

 Rearrangements occur in the neighborhood of several genes with interesting functions.  The Denisovan genome contains a non-duplicative structural rearrangement adjacent to \emph{BARD1}, a BRCA1 associated ring protein that functions as a DNA repair peptide and tumor suppressor.  Detrimental mutations in \emph{BARD1} inhibit the ability to perfom DNA repair and commonly result in widespread accumulation of chromosomal rearrangements, especially during tumor formation.  A second mutation in a gene with a NUDIX DNA repair domain is also observed, though the functional impacts (if any) of these mutations are unknown.  One rearrangement in the Denisova genome shuffles regions adjacent to two neuron-expressed genes \emph{CDK5RAP2} and \emph{CLOCK}.  \emph{CDK5RAP2} is a centrosomal protein with high expression in the brain \citep{Nagase2000}.  Deleterious mutations in \emph{CDK5RAP2} are associated with microcephaly and brain development abnormalities in humans\citep{Hassan2007, Bond2005,Pagnamenta2012, Lancaster2013}, while \emph{CLOCK} regulates circadian rhythm in model organisms \citep{darlington1998,vitaterna1994}.  \emph{CDK5RAP2} has experienced rapid amino acid substitutions in primates relative to rodents \citep{Evans2006}.   Both genes are known to be expressed in neurons and especially since this individual lived to adulthood there is no reason \emph{a priori} to suspect detrimental effects on gene functions. 

\subsection*{Rearrangements as potential barriers to gene flow}

Humans and Neanderthals interbred during range overlap in Eurasia, with an average of 1-2\% of Neanderthal DNA in modern European genomes \citep{Sankararaman2014,Green2010,Prufer2014}.  Yet, some portions of the human genome appear more resistant to introgression than others \citep{Sankararaman2014}.   Regions containing chromosomal rearrangements are less likely to experience introgression.  One potential explanation is that negative selection against new mutations prevents their spread from Neanderthals into modern humans.  However, variants that are derived in modern humans and ancestral in Neanderthals also exhibit a significant reduction in the likelihood of introgression.  Large chromosomal translocations are identified in spontaneous abortions \citep{DeBraekeleer1990,Goddijn2000,Fryns1998} and in individuals pursuing \emph{in vitro} fertilization treatments \citep{Schreurs2000}, consistent with a role for genome structure changes as barriers to reproduction.  Two potential genetic mechanisms can explain how regions housing chromosomal rearrangements that differentiate humans and Neanderthals might be associated with lower rates of introgression through effects of negative selection on F1 hybrids.  

\begin{figure}
\begin{center}
\includegraphics[scale=0.5]{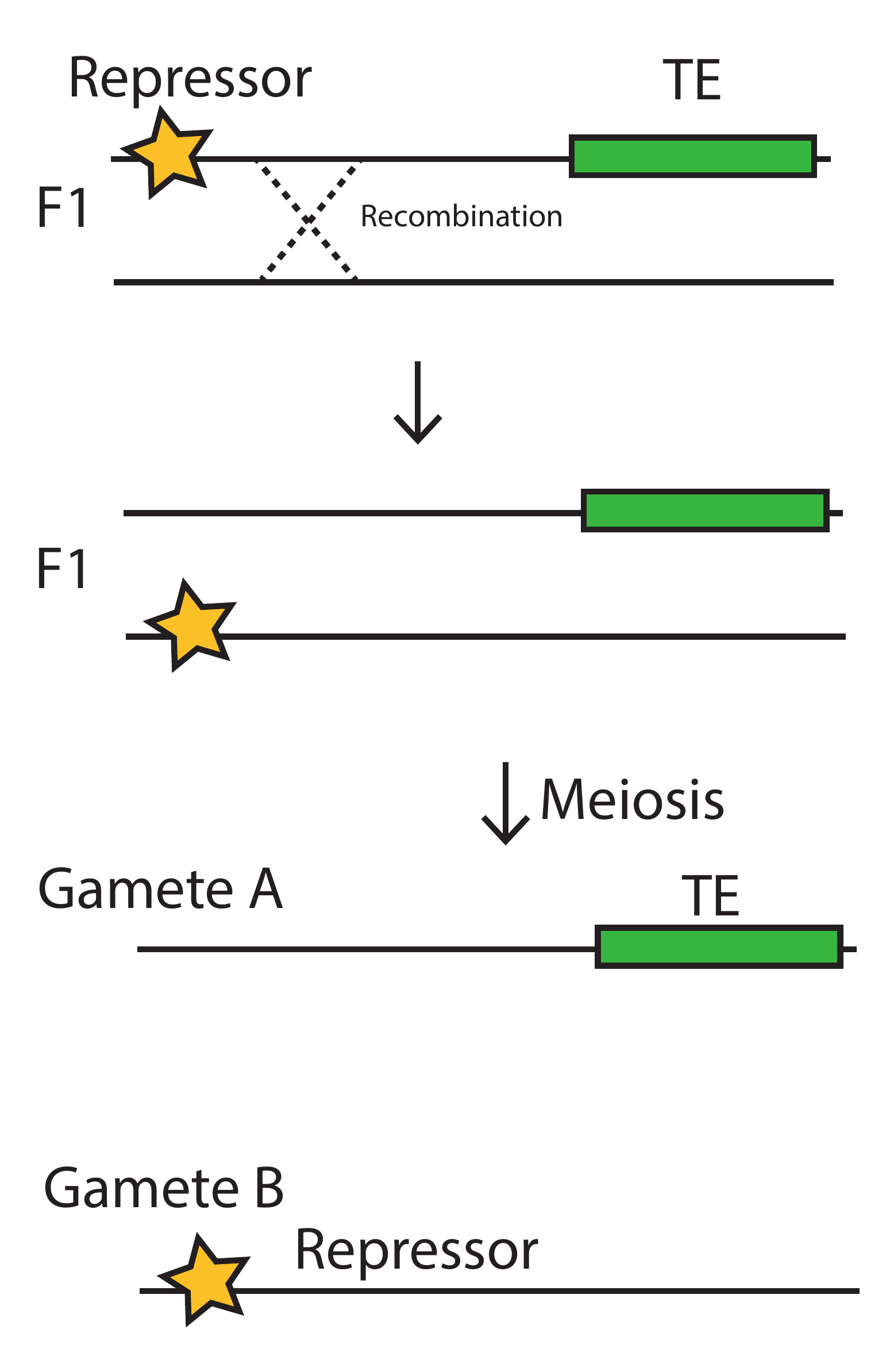}
\end{center}
\caption{\label{Repressor} Incompatible TE-repressor systems reduce hybrid fitness.  Transposable element-repressor systems become unlinked in a Neanderthal-human F1 hybrid.   Alternate segregation places the TE and repressor in separate gametes, inducing a TE burst in F2 offspring.  The detrimental effects of rampant TE movement would be expected to reduce fitness at the F2 generation.  TE bursts might also occur in F1s if repressor systems are sensitive to copy number and dosage.  Such incompatibilities could explain a portion of the reduced introgression observed.  } 
\end{figure}

First, translocations encompass multiple types of underlying mutations including gene conversion, ectopic recombination, retrogene formation, and transposable element mediated transposition.  It is likely that many of these mutations reflect TE activity or TE mediated recombination.   If different genomes contain incompatible TE-repressor systems, movement into new genetic backgrounds during interbreeding events could potentially incite activation of previously silenced TEs (Figure \ref{Repressor}).   TE activations are generally known to be deleterious and are a known source of reproductive incompatibilities in model organisms and plants \citep{Rubin1982, Petrov1995,Castillo2012} and similar molecular expansion of selfish elements has been observed in mammalian hybrids \citep{ONeil1998}.  If similar processes were to occur in humans they would explain a portion of the observed reduction in Neanderthal ancestry at regions housing changes in genome structure.  

\begin{figure}
\includegraphics[scale=0.32]{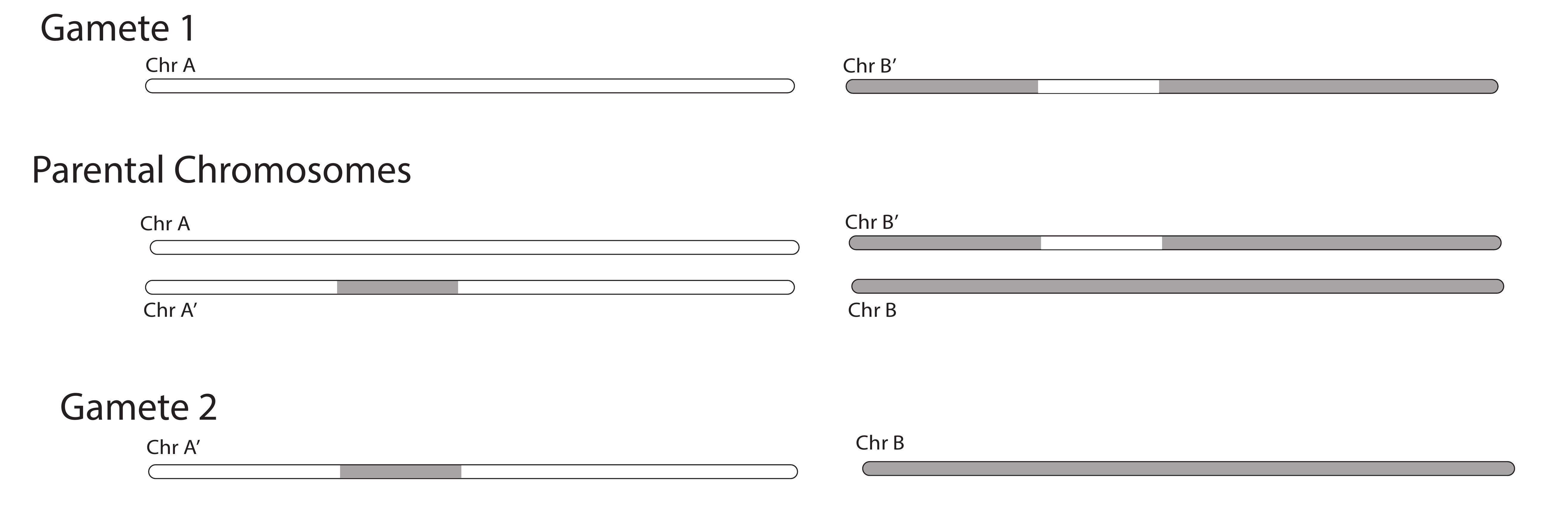}
\caption{\label{Meiosis} Segregation of rearrangement products produces incompatible chromosomes.  Chromosomal rearrangements A' and B', formed through reciprocal exchange of DNA across non-homologous chromosomes, undergo independent assortment during meiosis I.  If segregation of chromosomes is random, gametes have only a 50\% chance of inheriting only one rearranged chromosome.   Gametes will lack DNA captured by one rearrangement, but will contain additional copies of the complementary rearrangement segment.  If the rearrangement captures essential genes or regulatory factors necessary for development, parents with incompatible chromosomal rearrangements will have reduced fertility, and loss of non-essential genes can reduce offspring fitness.   Thus, even chromosomal rearrangements that do not have any other molecular or phenotypic effects when homozygous can reduce fitness in hemizygotes. } 
\end{figure}

Second, gain or loss of genes in gametes of  F1 offspring hemizygous for rearrangements might also reduce hybrid fitness \citep{Presgraves2010,Coyne2004} and contribute to the observed reduction in introgression from loci near chromosomal rearrangements (Figure \ref{Meiosis}).  Non-duplicative rearrangements form across chromosomes.  During meiosis in a hybrid individual, alternate segregation of chromosomal rearrangements places the mutant chromosomes in opposing daughter cells.  The resulting gametes contain duplicate copies of the rearrangement segment on one chromosome and a lack of the complementary rearrangement segment on the other chromosome.  If the rearrangement captures functional genes or regulatory elements necessary for survival or reproduction, offspring will have reduced fitness, resulting in barriers to genetic homogenization even in cases where rearrangements in and of themselves have no functional consequences in the F1 hybrid parent.    
 
Chromosomal rearrangements often do not spread through single populations due to lower lowered fitness in hemizygotes as well as potential negative molecular impacts of new mutations.  However, divergence for rearrangements could accumulate in allopatric separation, especially when aided by bottlenecks and inbreeding.  Modern humans experienced severe bottlenecks during the out of Africa migrations \citep{Li2011} resulting in lower genetic diversity for Eurasians.  Neanderthals experienced independent bottleneck events \citep{Prufer2014} allowing for accumulation of independent  mutations in the two groups.  Inbreeding in subpopulations can also spread accumulation of rearrangements as the associated decrease in heterozygosity could fix rearrangements in particular lineages \citep{Rieseberg2001}.  Neanderthals had low effective population sizes, low levels of heterozygosity, and instances of consanguineous mating \citep{Prufer2014}.  In the face of an F1 disadvantage described above, interbreeding after inbreeding would then be disfavored.  These factors together could contribute to differential accumulation of chromosomal rearrangements in archaic and modern humans, increasing the likelihood that they might later act as one potential barrier to genetic homogenization.    
 
\subsection*{Sex-specific changes}
 The X differs from the autosomes in dosage compensation \citep{Charlesworth1996}, sexual antagonism \citep{Rice1984}, dominance, recombination rate \citep{Schaffner2004}, and rate of amino acid substitution \citep{Mank2010}.  Similarly, the Y differs from the autosomes and the X in that it is only present in human males.  The Y has little recombination outside pseudoautosomal/XTR regions, reducing the efficiency of selection to sweep beneficial mutations to fixation and allowing greater potential for genetic hitchhiking \citep{Bachtrog2013}.  Chromosomal rearrangements with the Y can reduce male fertility \citep{Alves2002} even in cases of reciprocal translocation where genes are not gained or lost \citep{Dong2012}. Thus, changes in genomic locations from the sex chromosomes can result in alternate selective pressures, causing downstream changes in gene sequence evolution and gene expression \citep{Ellegren2007}.  Each of these factors could alter the evolutionary trajectories of relocated sequence, with important implications for sex chromosome evolution in humans.   

Beyond the typical molecular effects from changing genomic neighborhood, genomic trafficking across the sex chromosomes and autosomes therefore has the potential to alter selective pressures and selective constraints in ways that are not mimicked by autosome-autosome translocation. Using Illumina sequencing for female samples of Neanderthals and Denisovans, paired-end reads identify changes in genome structure that modify the sex specific status of surrounding DNA.   Thirty-three translocations exist between the X and autosomes in Neanderthals and 72 in Denisovans.   There is no excess of rearrangements on the X relative to its size.  It is possible that for the X negative selection against recessive deleterious mutations removing rearrangements from the X and that actual rates of formation are higher than those observed.  A total of 98 changes in genome structure map to the Y in Neanderthal and 110 map to the Y in Denisovans.  Each of these samples is female \citep{Prufer2014,Meyer2012}.  Yet, the data still capture genetic exchange between the Y and the autosomes or X, displaying the influence the Y can have on genome evolution in humans. Rates of rearrangements for small chromosomes are higher than for large chromosomes in both Denisovans and Neanderthals (Figure \ref{RecombRate}) and the Y shows large numbers of rearrangements relative to its size, consistent with this pattern.  The Y is degenerate, and commonly collects repetitive elements \citep{Bachtrog2013}.  Both the size of the Y and its association with selfish genetic elements may explain the large number of rearrangements on the Y.  However, given the lack of a Y in these female samples, the true rate of modification involving the Y will be even higher than observations presented here.

\subsection*{Testis biased expression}
Previous work in model organisms has shown that new genes are commonly expressed in the testes and later are exapted for alternative functions in other tissues \citep{Betran2002,Kaessmann2010,Assis2013}.   An excess of testis-specific genes affected by genome structure changes is found in Neanderthals compared to modern humans, but no such excess was observed in comparisons of Denisovans with modern humans.  Selective pressures in the testes commonly force rapid evolution in humans \citep{Wyckoff2000} and model organisms \citep{Haerty2007,Voolstra2007,Dorus2010}.  Additionally, large numbers of genes are testes-biased, and there may be permissive selective pressures that allow promiscuous expression in the testes.  Transposable elements that are active in the germline are more likely to be passed on in subsequent generations and these may also contribute to the observed association between testes expression and chromosomal rearrangements.  Testes-expressed genes show resistance to introgression between Neanderthals and modern humans across European populations \citep{Sankararaman2014}.  Here, chromosomal rearrangements are adjacent to testis-specific genes and loci with chromosomal rearrangements serve as one genetic factor that can create barriers to introgression due to negative selection on loci after interbreeding.  Thus, these results offer one specific genetic mechanism that may explain some portion of the observed association between testes specific expression and  barriers to interbreeding.  

\subsection*{Formation of a new gene sequence}
Among testis-specific genes associated with genome structure changes, one shows signals of particularly rapid evolution in genome architecture.  Multiple abnormally mapping read-pairs suggest rearrangement in the neighborhood of \emph{Fank1}, a nuclear protein expressed in sperm production during meiosis whose protein sequence is conserved across mammals \citep{Zheng2007}.  Knock-downs of \emph{Fank1} reduce fertility in mice by inducing apoptosis in developing sperm \citep{Dong2014}.   \emph{Fank1} also offers a rare example of allele-specific methylation in humans \citep{Li2010} and the gene displays elevated amino acid substitutions along the human lineage, indicative of positive selection.   The rearrangement of the first exon of \emph{Fank1} shows signs of 6-fold copy number variation in archaic genomes, as well as multiple breakpoints suggesting secondary modification.  \emph{Fank1} lies in a region with low recombination \citep{DeGiorgio2014}, and the disrupted synteny caused by the Y translocation can readily explain the observed lack of crossing over.  The region has been suggested to have  signals of balancing selection and ancient segregating polymorphism that matches exceptionally well with the region with identified copy number variation \citep{DeGiorgio2014}.  Given the precise match between genomic locations of  balancing selection signals and and coverage changes in modern and archaic human genomes, divergence of newly identified paralogs in the first exon and intron of \emph{Fank1} are likely to explain the unusual diversity patterns observed in balancing selection screens.  
 
A novel fusion transcript created by the rearrangement surrounding \emph{Fank1} now drives expression of a new exon in the testes for at least one modern human.   The region containing this new transcript has then experienced rampant copy number variation through secondary duplication with 6 copies in Neanderthals and 6 copies in Denisovan.  Combined with high rates of amino acid sequence evolution, this locus is subject to exceptionally rapid evolution in humans.  Thus, this verified case of genomic exchange between the Y and the autosomes as well as sperm-specific promoter derived from \emph{Fank1} makes this modified locus with new gene formation a strong candidate for functions in human reproduction.      

\section*{Methods}
\subsection*{Identifying structural variation}
All reads from the high quality 52X Altai Neanderthal genomic sequence (ERP002097) \citep{Prufer2014} and 38X Denisovan genomic sequence (kindly provided by Kay Pr\"ufer)  \citep{Meyer2012} were used to examine genome structure based on  abnormal mapping orientations.  These reads were aligned by the Neanderthal genome sequencing project against the human genome reference GRCh37 using bwa v.0.5.8a deactivating seeding and allowing for two gaps (options -l 16500 -n 0.01 -o 2) \citep{Meyer2012}.    A total of 84.4 million reads in Neanderthal and 132.7 million reads in Denisovans are long enough to generate non-overlapping sequences with independent alignments for read pairs, amounting to roughly 2.8X and 4.4X coverage that will be informative for genome structure.  Using samtools (\texttt{samtools view -f 1 -F 268}) I identified read-pairs where both partners mapped, considering only primary alignments.  Cases where reads mapped with a quality score $\geq 20$ where read-pairs aligned on different chromosomes or on the same chromosome at least 1 Mb apart identify  translocations. Variants supported by at least 3 but less than 100 read-pairs after clustering over a distance of 1000 bp were kept (Figure \ref{Support}).  

These methods cannot determine whether mutations are homozygous or heterozygous, and they may be limited in the face of identical repetitive elements.  However, rearrangements at many repetitive element sequences can be identified if they contain $\sim$1\% sequence divergence, allowing for distinguishable nucleotide sequences across the length of Illumina short reads.  Hundreds of transposable elements in the human reference genome contain such divergence \citep{Lander2001} and should be captured in these assays.  Repetitive sequence locations for all TEs provided by the UCSC genome browser for hg19/GRCh37 (downloaded June 2015) determined rearrangements whose breakpoints lie within selfish genetic elements.   To determine the number of rearrangements that might be consistent with duplicative rather than non-duplicative transfer of DNA, I searched for cases where at least one breakpoint had elevated coverage for 1 kb to the left or right of the abnormally mapping read pairs.  Sequence depth across the genome for reads mapping with a quality score $\geq 20$ was extracted using samtools (\texttt{samtools depth -q 20}).   Mean coverage for this region at or above a threshold of two standard deviations from the mean was considered to have elevated coverage.  Coverage is not always a reliable indicator of duplications \citep{Rogers2014}, and secondary mutations or artifacts of library preparation can also cause abnormal fluctuations in local coverage.  

Many rearrangements affect the Y, even though archaic samples are derived from female individuals \citep{Prufer2014,Meyer2012}, raising a possibility that false positives through mis-mapping might be driving results.   Five rearrangement variants in Neanderthal and 5 rearrangement variants in Denisovan that affect the Y outside pseudoautosomal regions  have a second blastn hit in the human reference genome within 2 kb of the autosomal or X read, each with between 91-99\% nucleotide identity.   These sites could potentially represent false positives due to mis-mapping of reads driven by allelic variation, or alternatively might be regions subject to homology mediated ectopic exchange especially through gene conversion events.  Thus, based on 5 out of 98 variants in Neanderthal and 5 out of 110 variants in Denisovan that might be the product of mis-mapping, the false positive rate for translocations involving the Y would lie between 0.00-5.10\% in Neanderthals and between 0.00-4.55\% in Denisova, consistent with false positive rates using paired-end read data to identify tandem duplications in model organisms \citep{Rogers2014}. 

Mean coverage per site across the entire Y is 0.906 for the Altai Neanderthal and 0.56 for the Denisovan while median coverage for the Y in both genomes is 0 (Table \ref{YChromCov}), very low coverage in comparison to male samples \citep{Prufer2014}.   In contrast, in Neanderthal regions within 1 kb of a translocation have a mean coverage of 74 while in Denisova regions within 1kb of rearrangement calls have a mean coverage of 52 (Table \ref{YChromCov}).  If Y contamination is uniform, heterogeneity in coverage between the whole Y compared to segments adjacent to rearrangements is not expected.  If Y rearrangements were driven by contamination from modern humans or from mis-mapping of reads, one would also expect to observe large numbers of within-chromosome translocations along the Y.   No within-chromosome translocations were identified in Neanderthals and only one  within-chromosome translocation affecting the Y in Denisovans.  This sole within-chromosome translocation lies adjacent to a region with a  translocation from the Y to an autosome, suggesting secondary rearrangement.  In addition to the observed translocations above, 22 cases of abnormal read-pair mapping in Neanderthal and 26 in Denisovan match to pseudoautosomal regions (PAR1 and PAR2) and X-translocated-region (XTR/PAR3) region of the X and Y.  These sites likely represent allelic variation that happens to match best with the human reference genome Y, in spite of actual location on the X.  These pseudoautosomal variants were therefore excluded from all downstream analyses as they are not indicative of translocations. 

\subsection*{Identifying ancestral states}

I used the gorilla genome as an outgroup to polarize autosomal sequences and the X. The chimpanzee genome is more closely related to humans, but was assembled relying on the  human genome scaffolds \citep{ChimpGenome}. In regions subject to genome structure changes, the chimpanzee genome commonly shows `N's indicating assembly uncertainty. The gorilla genome took advantage of technological advantages in next-generation sequencing and incorporated multiple sources of sequence data to resolve and order contigs \citep{GorillaGenome} and offers more reliable information.    The gorilla genome lacks a Y chromosome sequence, and therefore the chimpanzee genome was used as an outgroup for the Y to polarize the direction of mutations.  Autosomal and X mutations were polarized against the Gorilla reference genome r.3.1 provided by ENSEMBL.  Sequence matching with abnormally mapping read-pairs as well as 100 bp upstream and downstream was compared with all Gorilla chromosomes in a blastn at an E-value cutoff of $10^{-10}$.   For translocations mapping to the Y, a blastn search was used to match both breakpoints in the chimpanzee genome (r2.1.4).  Sequences matching with read-pairs mapped to the same chromosomal location within 1000 bp of one another were taken as cases where Neanderthals carry the ancestral state.  The ancestral state cannot be resolved for some mutations in cases where outgroups are incomplete or poorly assembled. 

\subsection*{Polymorphism in modern humans}
I confirmed mutations identified in archaic genomes using paired end read mapping for 10 modern human genomes collected and sequenced as part of the Neanderthal genome project (\url{http://cdna.eva.mpg.de/denisova/BAM/human/}, accessed Feb 2015, Table \ref{CellLines}) \citep{Meyer2012}.  Samples were prepared and sequenced in the same lab as part of the Neanderthal genome project and will be less likely to be subject to methodological differences than other human genomic samples (\url{http://cdna.eva.mpg.de/denisova/BAM/human/}, Accessed Feb 2015) \citep{Meyer2012}.  Samples are derived from immortalized cell lines, which commonly accumulate rearrangements.  Thus, I do not report the full genome wide structural variation for cell lines but rather focus on confirming mutations that are identified in archaic genomes.  

To determine whether rearrangements have accumulated in a manner consistent with constant neutral dynamics, I compared polymorphism to divergence for rearrangements and putatively neutral intergenic SNPs.  Methods used to identify rearrangements are unable to identify whether samples are heterozygous or homozygous in next-generation sequencing data, and they are unable to identify rearrangements except where the archaic genome and modern human reference genome differ.  To obtain an appropriate neutral comparison, I identified all SNPs where the archaic genome and the modern human reference genome differ for at least one allele, and then ascertained those SNPs in the the ten cell lines of modern humans, consistent with criteria used to identify rearrangement mutations.  Significance testing was performed using a chi-square test on the $2\times2$ contingency table of polymorphism and divergence for rearrangement mutations and neutral intergenic SNPs, similar to a McDonald-Kreitman test \citep{McDonald1991}.  

\subsection*{Gene expression in modern humans}
Genes whose transcription start or stop positions lie within 10kb of structural variants were identified, and evaluated for expression patterns against the human gene expression atlas.  Genes adjacent to structural variants were identified as ``expressed" if classified as ``medium" or ``high" in the Human Protein Atlas \citep{Fagerberg2014} (\url{http://www.proteinatlas.org/}, accessed Oct 2014).  10000 replicates of equal numbers of genes were chosen to determine the likelihood that as many genes would have tissue specific expression.  Genes with testis specific changes in expression between humans and chimpanzees were taken from \cite{Khaitovich2005}.  Gene annotations from NCBI (\url{ftp://ftp.ncbi.nlm.nih.gov/gene/DATA/gene_info.gz}, Accessed Jan 25 2015)  and DAVID gene ID converter (\url{http://david.abcc.ncifcrf.gov/}, Accessed Jan 25 2015) matched previous annotations with ENSEMBL gene identifiers.  Out of 11,780 genes, 9,083 genes had identifiers that matched with current annotations in ENSEMBL.  Overrepresentation of testes-specific expression patterns was identified by resampling 10,000 replicate datasets of randomly sampled genes of the same size as that observed.  

Transcriptome data for testes of a 44 year old male modern human individual from the ENCODE project, provided by Michael Snyder's lab (accession ENCSR693GGB, \url{www.encodeproject.org}, accessed March 2015) was used to validate  new gene formation in the region of \emph{Fank1}. I used tophat-fusion search \citep{Trapnell2009,Kim2011} to map fastq reads to all major chromosomes for the human reference genome GRCh 37.75. I used tophat v.2.0.13, with command line options \texttt{--fusion-search --fusion-min-dist 1000000 --fusion-read-mismatches 4}, and all other parameters set to default.  Tophat was run using bowtie2 v2.2.5.  

\subsection*{Rates of Introgression}

Introgression data from Sankararaman et al. 2014 and from Steinr\"ucken et al. 2015 (\url{http://dical-admix.sf.net})  was used to establish whether regions with chromosomal rearrangements were less likely to experience introgression.   Steinr\"ucken et al. offer probabilistic calls by window for 500 bp windows across the genome.  Mean probability of introgression per site per strain for chromosomal rearrangements was compared with random resampling datasets to establish the probability of observing results as low or lower than random expectations.  Resampling estimates used  10,000 replicates, choosing windows at random throughout the genome.   Windows on the X and Y were excluded.  The Y does not have introgression data because samples are female and the X is subject to lower levels of introgression for reasons potentially unrelated to genome structure that might influence results.  Sankararaman et al's data offers calls for sites that experienced introgression. The proportion of sites with introgression tracts in at least one haplotype for regions containing genome structure calls was compared with probability of success set to the background rate of 35.64\% of the genome using a binomial test.  

\subsection*{Rapidly evolving genes}
A reciprocal-best-hit blastn search at an E-value cutoff of $10^{-10}$ defined orthologs for all CDS annotations for \emph{Gorilla gorilla} r.3.1, \emph{Pan troglodytes} r2.1.4, and \emph{Homo sapiens} GRCh 37.75.  Genes with clear one to one ortholog calls across Human-Chimp-Gorilla were then used for further analysis.  Protein sequences for genes were aligned using clustalw 2.1 \citep{Larkin2007} and back-translated protein alignments to generate in-frame nucleotide alignments.  $d_N/d_S$ was estimated in PAML using the F1x4 a codon model which estimates codon frequencies based on nucleotide frequencies, estimating $\kappa$ with an initial $\kappa=2.0$.  
 
\subsection*{PacBio confirmation of structural variation at the \emph{Fank1 Locus}}
Targeted variants were then confirmed in long molecule sequencing collected from a haploid complete hydatidiform mole provided by Pacific Biosciences (\url{http://datasets.pacb.com/2014/Human54x/fast.html}, Accessed March 2015) recently used to generate a \emph{de novo} human genome assembly \citep{Steinberg2014}. Reads were aligned to major chromosomes for the human genome reference GRCh37.75 using blasr \citep{Chaisson2012}, reporting the best 10 matches (\texttt{-bestn 10}) and all other parameters set to default.   The PacBio aligner blasr favors long alignments and often does not report shorter split-read alignments even those with greater nucleotide similarity.  All PacBio sequence reads that matched to the location that contains rearrangement signals in Illumina sequences for the targeted site were aligned using a blastn search against the human reference genome at a cutoff of $E\leq10^{-10}$ similar to methods that have successfully confirmed genome structure variation \citep{Rogers2014}. Split read mappings that align for 1 kb or more on either side of the breakpoint defined by Illumina sequencing read-pair data confirmed this translocation.  Confirming reads also match the expected orientation indicated by Illumina sequencing reads.

\subsection*{PacBio confirmation of high frequency variants}
Some rearrangement variants identified in Neanderthals were also identified in 10/10 modern human genome samples in the HGDP panels.  To ascertain that these high frequency variants were not the product of bioinformatic artifacts, I confirmed these variants using a subset of PacBio long molecule sequence data.  Fasta files were downloaded from Pacific Biosciences (\url{http://datasets.pacb.com/2014/Human54x/fast.html}, Accessed March 2015) and aligned using a BLASTn search at an E-value of $10^{-10}$ against a reduced reference database comprised of genomic segments including 20 kb upstream and 20 kb downstream of high frequency rearrangements.  This reduced reference database was essential to make confirmation computationally tractable on a genome wide scale.  I then searched for single reads that produced alignments at least 200 bp long on each side of the rearrangement breakpoints within a span of 1000 bp of the rearrangement breakpoint.  

\subsection*{Acknowledgements}
I would like to thank Kelley Harris, Qi Zhou, and Montgomery Slatkin for helpful input concerning analyses and helpful comments on the manuscript.  Thanks also to Kay Pr\"ufer for sending unfiltered bam files for the Denisova analysis.  Matthias Steinr\"ucken and Yun Song graciously shared data on Neanderthal introgression prior to publication, which was essential for analyses presented in the current work.  RLR is funded by NIH grant R01-GM40282 to Montgomery Slatkin.

\clearpage
\bibliographystyle{MBE}
\bibliography{NeanderthalGenomeStruct}

\clearpage

\begin{threeparttable}
\begin{center}
\caption{\label{Mutations} Genome structure changes identified in Neanderthals and Denisovans }

\begin{tabular}{lcc}
\hline
Type & Neanderthal & Denisovan \\
Duplicative$^1$ & 212 & 327 \\
Derived$^2$ & 287 & 539 \\
\hline
Transposition &  131 & 236 \\
TE Ectopic Exchange &315 & 543   \\
Non-TE & 539   & 551 \\
\hline
Derived$^3$ & 336  & 599  \\
Ancestral$^3$ & 348 & 357 \\
Unknown$^3$ & 301 &  374 \\
\hline
Total & 985  & 1330 \\
\hline
\end{tabular}
\begin{tablenotes}
\item[1] Adjacent to region with coverage two standard deviations above mean genomic coverage. 
\item[2] Mutations on autosomes or X known to be derived in the archaic genome where coverage can be assayed.
\item[3] Mutations polarized against gorilla (X and autosomes) or chimpanzee (Y chromosome).  Some cannot be successfully polarized due to poor assembly of outgroups.  
\end{tablenotes}
\end{center}
\end{threeparttable}

\clearpage
\begin{center}
\begin{threeparttable}
\caption{\label{Divergence} Polymorphism and Divergence for Rearrangements$^1$ and SNPs }

\begin{tabular}{lrr}
 & Neanderthal & Denisovan \\ 
\hline
Polymorphic Rearrangements &  504   & 494 \\
Divergent Rearrangements &  383 & 725  \\ 
\hline
Polymorphic SNPs$^2$ & 161898 & 1092087\\
Divergent SNPs$^2$ & 130476 & 986390 \\ 
\hline
\end{tabular}
\begin{tablenotes}
\item[1] Excludes Y chromosome variants, which lack a SNP comparison for archaic humans.  
\item[2] Intergenic SNPs for neutral comparison. 
\end{tablenotes}
\end{threeparttable}
\end{center}
\clearpage

\begin{threeparttable}
\begin{center}
\footnotesize
\caption{\label{Testes} Overrepresentation of testes-specific expression patterns }

\begin{tabular}{lccccc} 
Genome  & testis-specific$^1$ & & chimp-human testis diverged$^2$   \\
\hline
Altai Neanderthal &   15/124 & $P=0.0084$  & 9/25 & $P=0.019$  & Fisher's$^3$ $P=0.0014$ \\
Denisovan   & 14/182   & $P=0.1422$ &   10/45 & $P=0.321$ & Fisher's $P=0.18$\\
\hline
\end{tabular}
\begin{tablenotes}
\item[1] Genes with testis-specific expression in modern humans.  Data from Human Protein Atlas. 
\item[2] Genes with testis-specific changes between humans and chimpanzees. Data from Khaitovich et al. 2005. 
\item[3] Fisher's combined P-value for the two tests of testis-specific association.
\end{tablenotes}
\end{center}
\end{threeparttable}

\clearpage

\begin{table}
\begin{center}
\small
\caption{\label{Introgression} Introgression rates from Neanderthal into modern humans }

\begin{tabular}{lccccc}
 Type &introgression probability & genomic background & standard deviation &  $P$-value   \\
\hline
All structural & 0.008   & 0.012  & 0.00083 &   $\leq10^{-3}$  \\
Derived in modern humans  & 0.007 & 0.012 & 0.00083 &   0.0015  \\
\hline
\end{tabular}
\end{center}
\end{table}

\clearpage

\clearpage{}

\clearpage{}

\clearpage{}

\clearpage

\clearpage{}

\clearpage{}

\clearpage{}


\section*{Supplementary Information}
\renewcommand{\thefigure}{S\arabic{figure}}
\renewcommand{\thetable}{S\arabic{table}}
\setcounter{figure}{0}
\setcounter{table}{0}
\setcounter{page}{1}

\newpage

\begin{table}
\caption{\label{ByChrom}Rearrangements by chromosome}
\begin{center}
\footnotesize
\begin{tabular}{llc}
\hline
Neanderthal & 1 & 122 \\
& 2 & 172 \\
& 3 & 61 \\
& 4 & 111 \\
& 5 & 31 \\
& 6 & 19 \\
& 7 & 202 \\
& 8 & 27 \\
& 9 & 122 \\
& 10 & 132 \\
& 11 & 66 \\
& 12 & 22 \\
& 13 & 48 \\
& 14 & 31 \\
& 15 & 69 \\
& 16 & 145 \\
& 17 & 133 \\
& 18 & 44 \\
& 19 & 36 \\
& 20 & 110 \\
& 21 & 107 \\
& 22 & 27 \\
& X & 35 \\
& Y & 98 \\
\hline
Denisovan & 1 & 128 \\
& 2 & 252 \\
& 3 & 99 \\
& 4 & 180 \\
& 5 & 36 \\
& 6 & 43 \\
& 7 & 283 \\
& 8 & 39 \\
& 9 & 119 \\
& 10 & 233 \\
& 11 & 82 \\
& 12 & 58 \\
& 13 & 28 \\
& 14 & 37 \\
& 15 & 67 \\
& 16 & 215 \\
& 17 & 124 \\
& 18 & 73 \\
& 19 & 66 \\
& 20 & 132 \\
& 21 & 136 \\
& 22 & 39 \\
& X & 79 \\
& Y & 112 \\
\hline
\end{tabular}
\end{center}
\end{table}
\clearpage

\begin{table}
\begin{center}
\caption{\label{GO} Overrepresented Gene Ontology Categories }

\begin{tabular}{lccc}
Genome  &  function & EASE    \\
\hline
Altai Neanderthal & flotillin  & 1.27  \\
Denisovan   & keratin &  1.28 \\
& microtubule & 1.17 \\
	&  flotillin & 1.14 \\
	& fibronectin & 1.13 \\	
\hline
\end{tabular}
\end{center}
\end{table}

\clearpage

%
%
%

\clearpage
\begin{table}
\caption{\label{PairCov} Mean coverage in properly paired reads by chromosome}
\begin{center}
\footnotesize
\begin{tabular}{llc}
\hline
Neanderthal &1 & 3.03442 \\
&2 & 3.11308\\
& 3 & 3.072\\
& 4 & 3.21629\\
& 5 & 3.11201\\
& 6 & 3.17608\\
& 7 & 3.09368\\
& 8 & 3.17866\\
& 9 & 3.00747\\
& 10 & 3.17219\\
& 11 & 2.96935\\
& 12 & 2.99443\\
& 13 & 3.16641\\
& 14 & 3.02113\\
& 15 & 2.93645\\
& 16 & 2.99368\\
& 17 & 2.77537\\
& 18 & 3.08382\\
& 19 & 2.58384\\
& 20 & 2.78167\\
& 21 & 3.23365\\
& 22 & 2.56239\\
& X & 3.16043\\
\hline
Denisovan & 1 & 3.03442 \\
& 2 & 3.11308 \\
& 3 & 3.072 \\
& 4 & 3.21629 \\
& 5 & 3.11201 \\
& 6 & 3.17608 \\
& 7 & 3.09368 \\
& 8 & 3.17866 \\
& 9 & 3.00747 \\
& 10 & 3.17219 \\
& 11 & 2.96935 \\
& 12 & 2.99443 \\
& 13 & 3.16641 \\
& 14 & 3.02113 \\
& 15 & 2.93645 \\
& 16 & 2.99368 \\
& 17 & 2.77537 \\
& 18 & 3.08382 \\
& 19 & 2.58384 \\
& 20 & 2.78167 \\
& 21 & 3.23365 \\
& 22 & 2.56239 \\
& X & 3.16043 \\
\hline
\end{tabular}
\end{center}
\end{table}

\begin{table}
\begin{center}
\caption{\label{YChromCov} Y Coverage}
\small
\begin{tabular}{llccc}
Genome & Locus & Mean Coverage  & Median Coverage & Standard Deviation \\
\hline
Neanderthal & whole  Y & 0.91 & 0 &  18.5 \\
 & within 1kb of Y-translocations &  74 & 14 & 220 \\
\hline
Denisovan & whole Y & 0.56  & 0 &  12.5 \\
&  within 1kb of Y-translocation &  52 & 7 & 160 \\
\hline

\end{tabular}
\end{center}
\end{table}

\clearpage

\clearpage
\begin{table}
\caption{\label{CellLines}Human immortalized cell lines used to confirm rearrangements}
\begin{center}
\footnotesize
\begin{tabular}{l}
Cell line \\
\hline
HGDP00521 \\ 
HGDP00542  \\
HGDP00665 \\
HGDP00778 \\
HGDP00927 \\
HGDP00998 \\
HGDP01029 \\
HGDP01284 \\
HGDP01307 \\
HGDP0456 \\
\hline
\end{tabular}
\end{center}
\end{table}

\clearpage
\begin{figure}
\begin{subfigure}{.5\textwidth}
\includegraphics[scale=0.47]{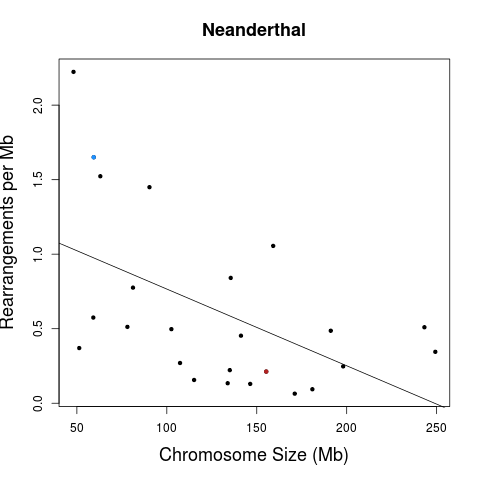}
\caption{}
\end{subfigure}
\begin{subfigure}{.5\textwidth}
\includegraphics[scale=0.47]{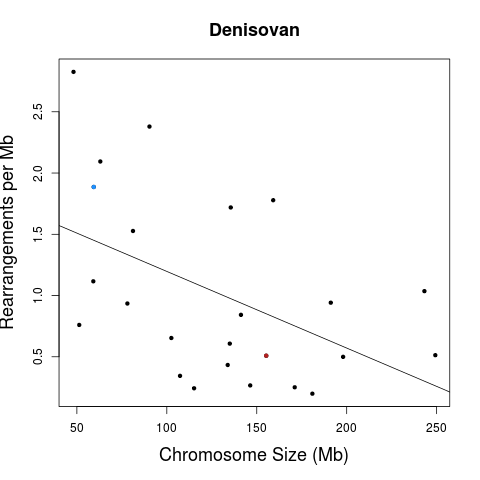}
\caption{}
\end{subfigure}
\caption{\label{RearrLength} Incidence of rearrangements vs chromosome size for (A) Neanderthal and (B) Denisova.   Both samples show a significant negative correlation between the rate of rearrangement and chromosome size (Neanderthal $R^2=0.24$, $P=0.0088$; Denisovan $R^2=0.20$, $P=0.016$).    These results may suggest that ectopic recombination is more common for small chromosomes or that small chromosomes are degenerating.   Rates of rearrangements are higher in Denisovan than Neanderthal, but  chromosomes show consistent patterns across the two species.  The X chromosome (red) does not appear to have an excess of rearrangements given its size, but the Y chromosome (blue) carries a large number of rearrangements per basepair, especially considering that there is no Y chromosome sequence present in these female individuals.}
\end{figure}
\clearpage{}

\begin{figure}

\begin{subfigure}{.5\textwidth}
\includegraphics[scale=0.45]{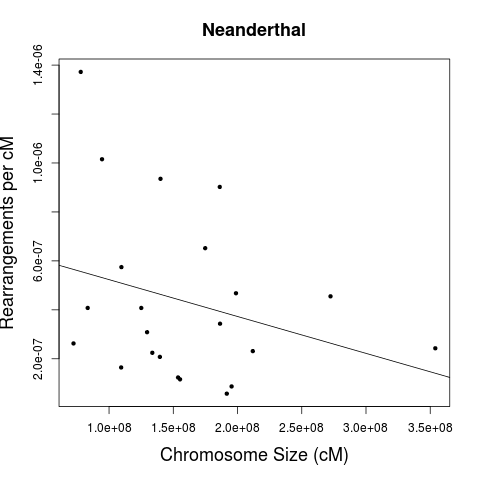}
\caption{}
\end{subfigure}
\begin{subfigure}{.5\textwidth}
\includegraphics[scale=0.45]{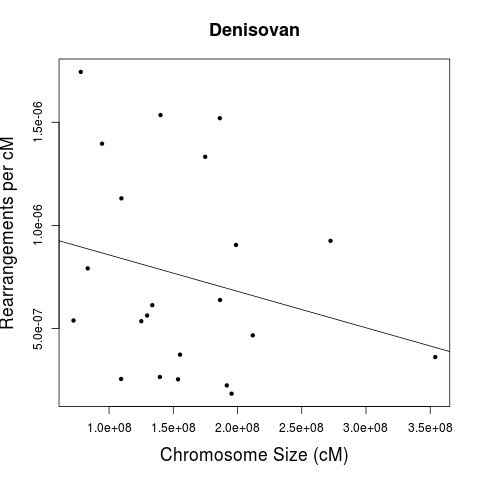}
\caption{}
\end{subfigure}
\caption{\label{RecombLen} Rate of rearrangements vs chromosome size in centimorgans for (A) Neanderthal and (B) Denisova.   There is no significant correlation between the rate of rearrangement and chromosome size (Neanderthal $R^2=0.036$, $P=0.195$; Denisovan $R^2=0.01$, $P=0.28$).}
\end{figure}
\clearpage{}

\begin{figure}
\begin{subfigure}{.5\textwidth}
\includegraphics[scale=0.47]{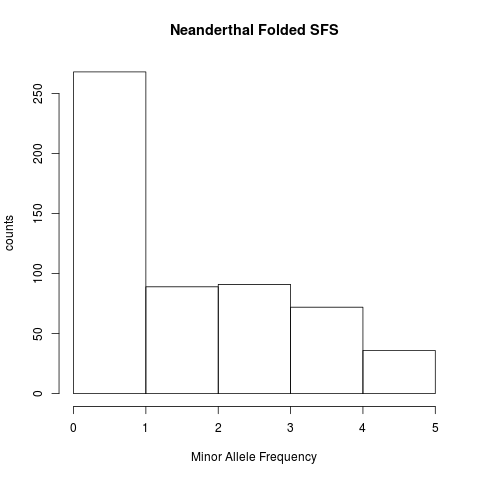}
\caption{}
\end{subfigure}
\begin{subfigure}{.5\textwidth}
\includegraphics[scale=0.47]{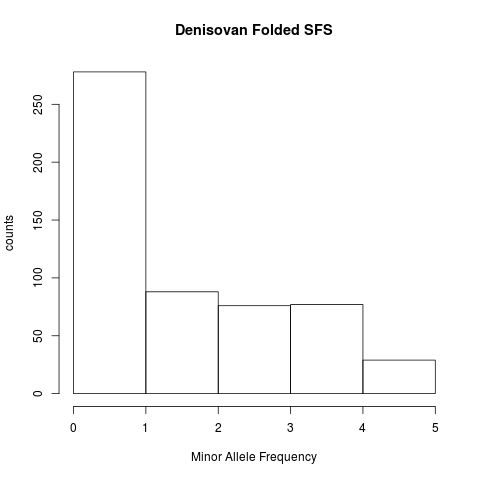}
\caption{}
\end{subfigure}
\caption{\label{SFS} Folded presence-absence spectrum for genome structure changes identified in (A) Neanderthal and (B) Denisovan genomes assayed in a population of 10 individual modern humans.  The presence-absence spectrum indicates a skew toward high and low frequency variants with fewer moderate frequency variants.  There is no significant difference between the folded mutation spectrum for rearrangements identified in Neanderthals vs. Denisovan.  }
\end{figure}
\clearpage{}

\begin{figure}
\begin{subfigure}{.5\textwidth}
\includegraphics[scale=0.5]{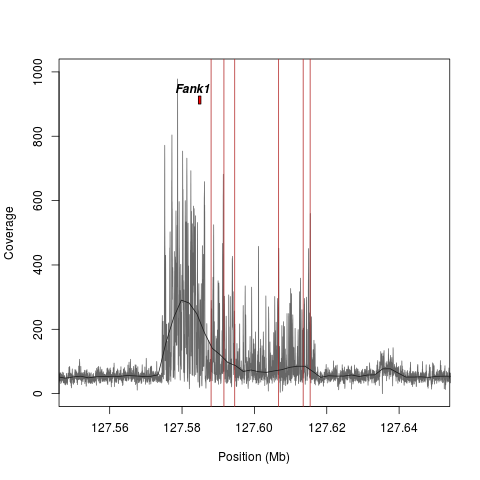}
\caption{}
\end{subfigure}
\begin{subfigure}{.5\textwidth}
\includegraphics[scale=0.5]{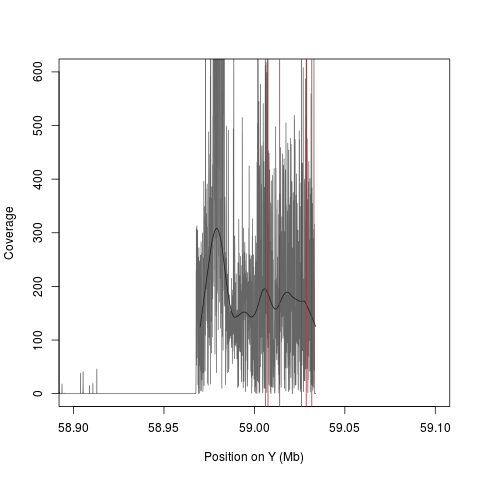}
\caption{}
\end{subfigure}
\caption{\label{FankDepth}  Genomic coverage with lowess smoothed regression line at the site of a rearrangement (A) at the \emph{Fank1} locus and (B) on the translocated segment of the Y in the Altai Neanderthal.  Locations of abnormally mapping read pairs that indicate junctions of rearrangements are shown in red.  Coverage changes are consistent with 6 fold copy number variation, and coverage depth is variable across the region, consistent with multiple independent breakpoints indicated by the paired end read information.  } 
\end{figure}

\clearpage{}

\begin{figure}
\begin{subfigure}{.5\textwidth}
\includegraphics[scale=0.5]{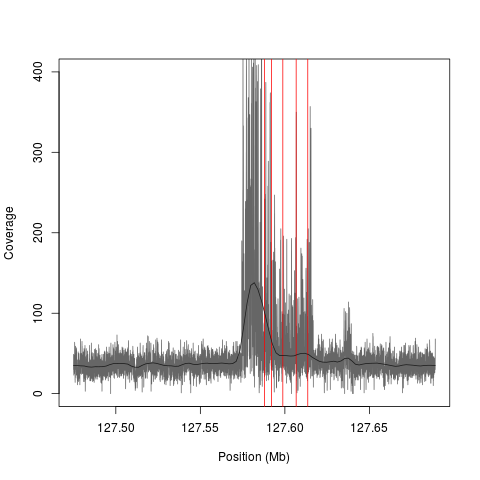}
\caption{}
\end{subfigure}
\begin{subfigure}{.5\textwidth}
\includegraphics[scale=0.5]{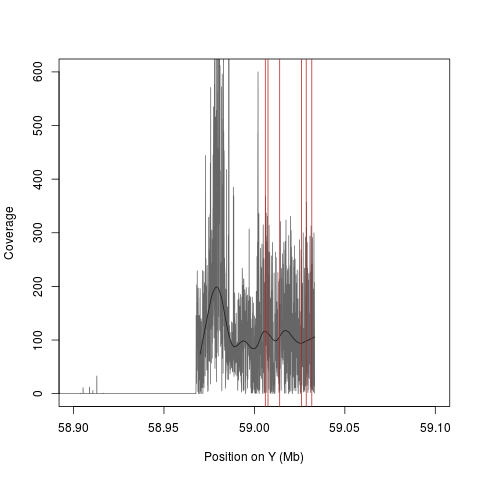}
\caption{}
\end{subfigure}
\caption{\label{DeniFankDepth}  Genomic coverage with lowess smoothed regression line at the site of a rearrangement (A) at the \emph{Fank1}locus and (B) on the translocated segment of the Y in Denisovan.  Locations of abnormally mapping read pairs that indicate junctions of rearrangements are shown in red. } 
\end{figure}

\clearpage{}

\begin{figure}
\begin{subfigure}{.5\textwidth}
\includegraphics[scale=0.5]{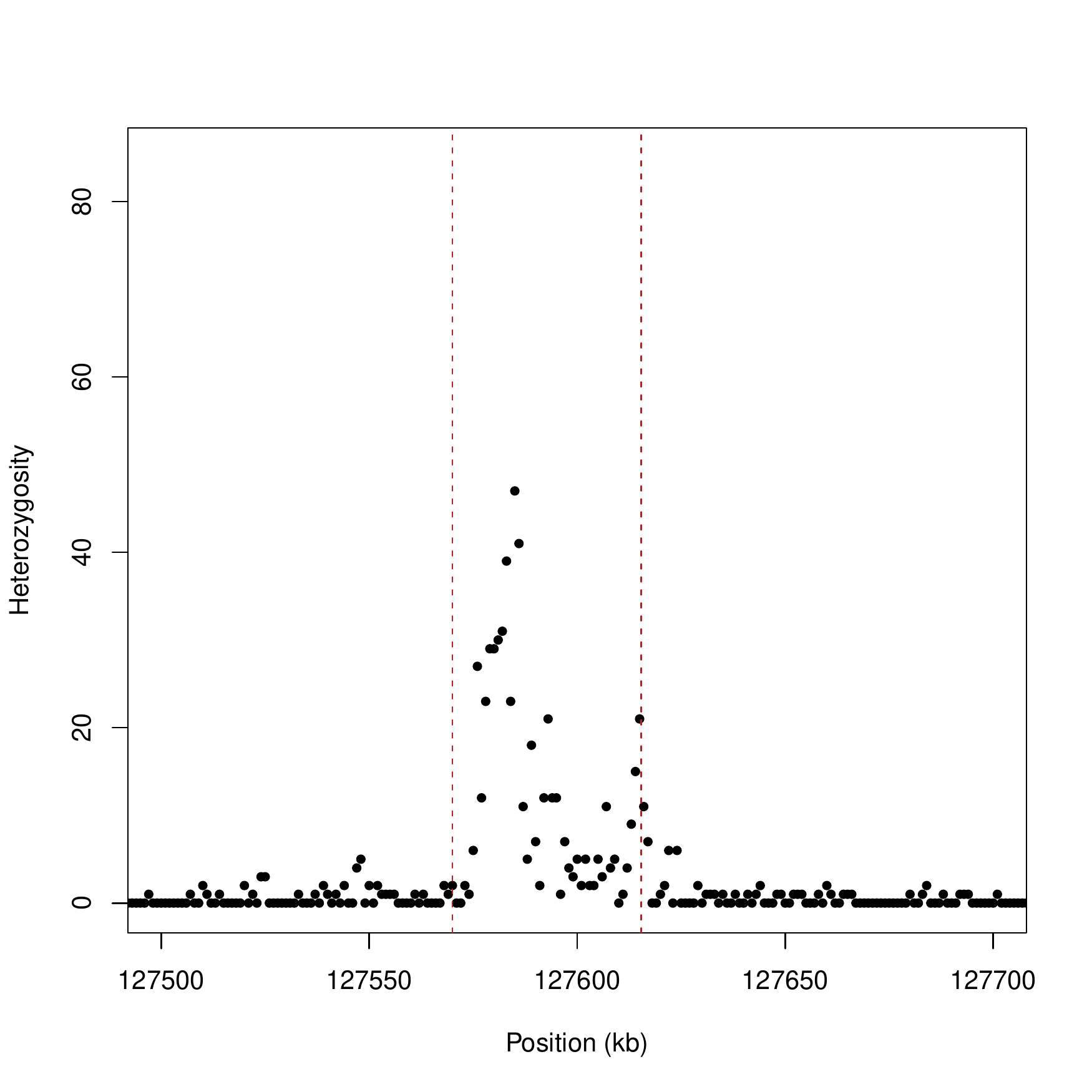}
\caption{}
\end{subfigure}
\begin{subfigure}{.5\textwidth}
\includegraphics[scale=0.5]{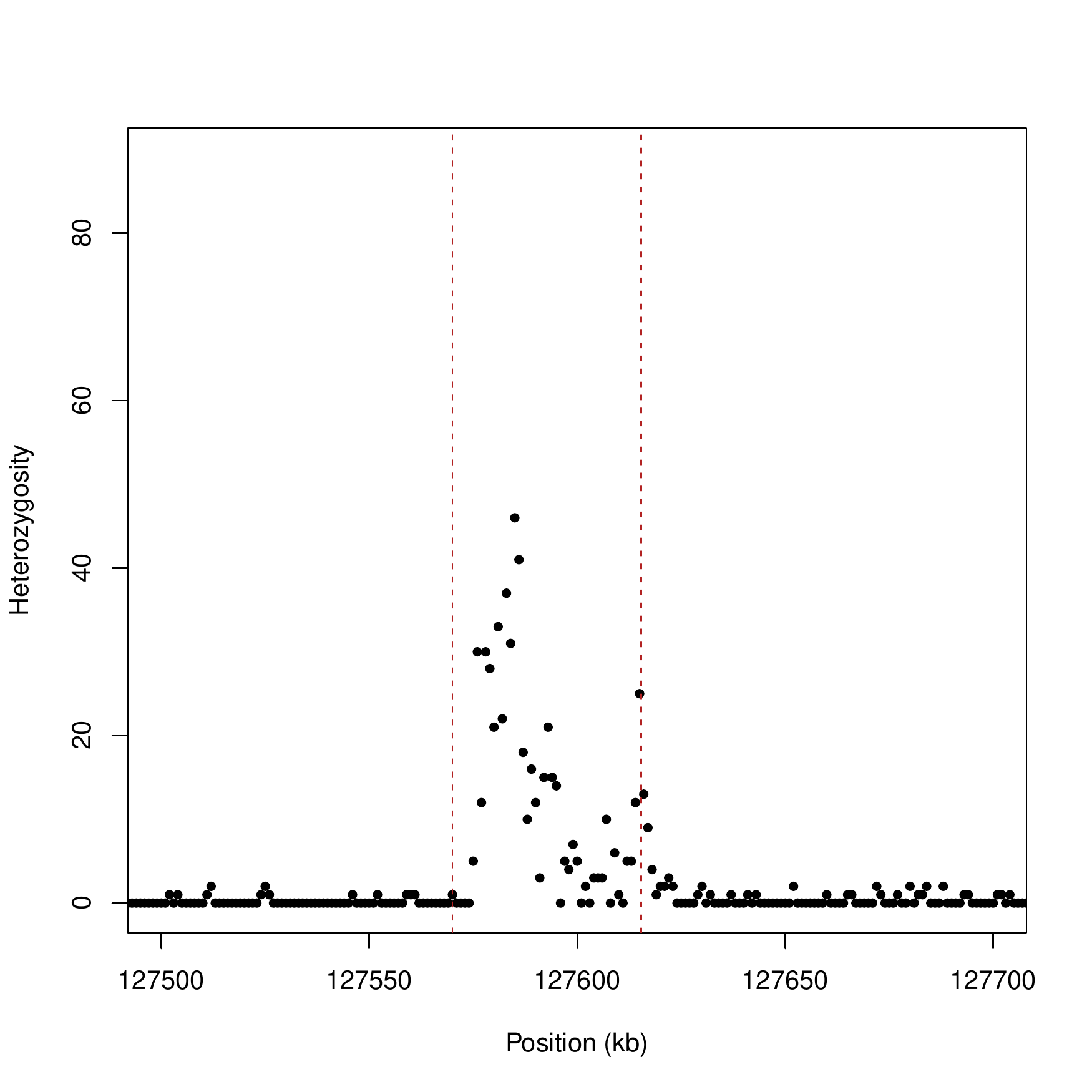}
\caption{}
\end{subfigure}
\caption{\label{FankHet}  Heterozygosity for 1 kb windows surrounding the duplicated first exon of \emph{Fank1} in (A) Neanderthal and (B) Denisovan.  Boundaries of the duplication inferred from coverage data and abnormally mapping read pairs is shown in red.   Heterozygosity for the region is abnormally high, a signature of paralogs accumulating mutations and diverging over time.  Heterozygosity is highest in the regions with higher copy number status and return to normal levels outside the duplicated region in \emph{Fank1}. } 
\end{figure}

\clearpage{}

\begin{figure}
\includegraphics[scale=0.9]{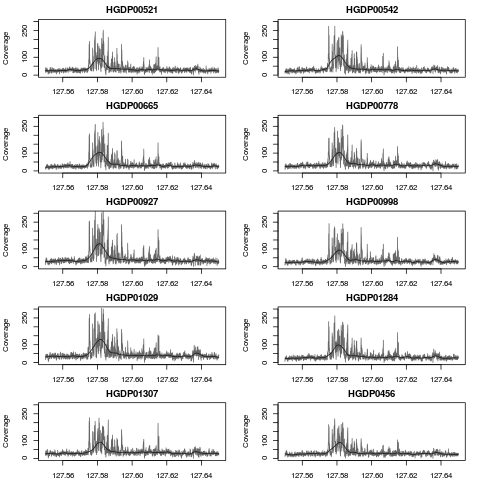}
\caption{\label{HGDPCov} Genomic coverage depth for the translocated segment of the chromosome 10 for 10 modern human genomes.  Modern humans show increased coverage consistent with multiple copies for the region.  } \end{figure}

\clearpage{}

\begin{figure}
\includegraphics[scale=0.9]{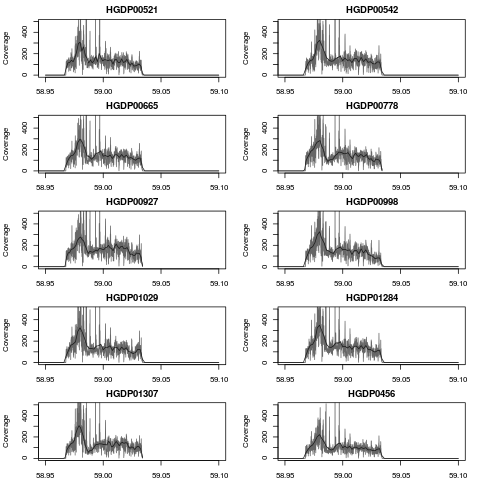}
\caption{\label{HGDPYCov} Genomic coverage depth for the translocated segment of the Y in 10 modern human genomes.  Modern humans show increased coverage consistent with multiple copies for the region.  } \end{figure}

\clearpage{}

\clearpage
\begin{figure}
\begin{subfigure}{.5\textwidth}
\includegraphics[scale=0.47]{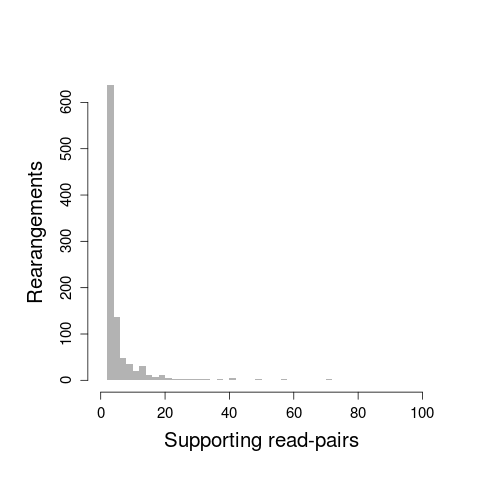}
\caption{}
\end{subfigure}
\begin{subfigure}{.5\textwidth}
\includegraphics[scale=0.47]{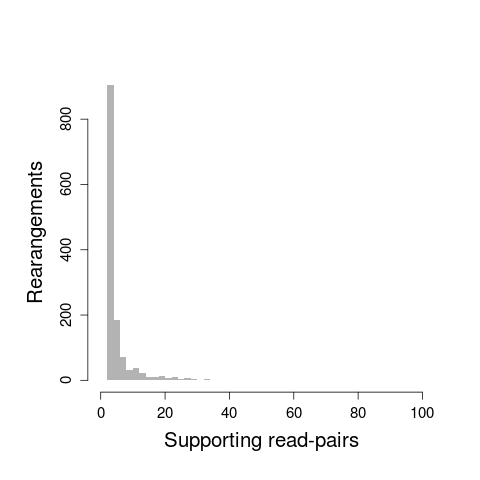}
\caption{}
\end{subfigure}
\caption{\label{Support} Read pairs supporting genome structure changes identified in (A) Neanderthal and (B) Denisovan genomes. }
\end{figure}
\clearpage{}
\end{document}